\pdfoutput=1
\PassOptionsToPackage{table, xcdraw}{xcolor}
\documentclass[PAPER, cernpreprint, atlasdraft=false, UKenglish, texmf, orcidlogo]{atlasdoc}
\usepackage[section]{placeins}
\usepackage{enumitem}

\usepackage[full]{atlaspackage}
\usepackage{atlasbiblatex}

\usepackage[jetetmiss=true, process=true]{atlasphysics}
\graphicspath{{logos/}{figures/}}

\usepackage{ANA-HMBS-2024-24-PAPER-defs}

\xspaceaddexceptions{",), }

\AtlasTitle{Search for Lorentz-boosted di-$\tau$ resonances produced in association with top quark pairs in $\sqrt{s}=13\ $TeV pp collisions with the ATLAS detector}

\AtlasAbstract{%
A search for a Lorentz-boosted pseudoscalar resonance $a$ in the mass range between $20\ $GeV and $85\ $GeV, produced in association with a top-antitop quark pair and decaying into a $\tau$-lepton pair, is presented. Both $\tau$-leptons are reconstructed in the hadronic decay mode. The boosted production of $a$ results in a small angular separation between its decay products, requiring the use of dedicated boosted di-$\tau$ objects with custom reconstruction, energy calibration, and identification procedures. The analysis uses proton--proton collision data at a centre-of-mass-energy of $\sqrt{s}=13\ $TeV collected with the ATLAS detector at the Large Hadron Collider, corresponding to an integrated luminosity of 140 fb$^{-1}$.
No significant deviation is observed over the expected Standard Model background, and a model independent upper limit of 0.19~fb is placed at the 95\% confidence level on the visible cross-section. Limits on $\sigma(t\bar{t}a) \times \text{BR}(a \to \tau\tau)$ are also obtained for a Two-Higgs-Doublet Model signal scenario, ranging from 0.4~pb at $m_a=20\ $GeV to 0.05~pb at $m_a=85\ $GeV.
}

\AtlasRefCode{HMBS-2024-24}

\PreprintIdNumber{CERN-EP-2026-205}

\AtlasJournal{Journal of High Energy Physics}
\AtlasCoverEgroupAnalysisTeam{atlas-hmbs-2024-24-analysis-team@cern.ch}

\hypersetup{pdftitle={ATLAS document},pdfauthor={The ATLAS Collaboration}}

\newcommand{\asymcol}[4]{ \multicolumn{1}{S@{\(\,^{+}_{-}\,\)}}{#1} & \multicolumn{1}{l}{\hspace{#4}$^{#2}_{#3}$} }

\begin{document}

\maketitle

\tableofcontents

\section{Introduction}
%

% The next lines are included from the .//chapters/introduction.tex input file
The discovery of the Higgs boson in 2012 by the ATLAS~\cite{HIGG-2012-27} and CMS~\cite{CMS-HIG-12-028} collaborations at the Large Hadron Collider (LHC)~\cite{Evans:2008zzb} established the Standard Model (SM) mechanism for electroweak symmetry breaking, but it did not exclude the possibility of an extended Higgs Sector. Many scenarios of physics beyond the Standard Model (BSM) postulate such extensions~\cite{Lee:1973iz,Dobrescu:2000yn,Ellwanger:2003jt,Branco:2011iw,Wang2012}, and their parameter spaces generally allow for the presence of Higgs-like states with masses below 125 \GeV. Such light bosons arise in a variety of contexts, including Dark Matter models~\cite{Silveira:1985rk,Pospelov:2007mp,Hektor:2014kga,Boehm:2014hva,Ipek:2014gua,Arina:2014yna}, neutral naturalness scenarios~\cite{Chacko:2005pe,Burdman:2014zta,Craig:2015pha}, models with a first-order electroweak phase transition~\cite{Profumo:2007wc,Blinov:2015sna,Ghorbani:2018yfr}, composite-Higgs models~\cite{Cacciapaglia:2019bqz}, and dilaton~\cite{Ahmed:2019csf} and axion-like particle~\cite{Mimasu:2014nea,Bauer:2017ris,Brivio:2017ije} scenarios.

Searches for light Higgs-like particles in association with top--antitop (\ttbar) quark pairs provide an important probe of scenarios in which such (pseudo)scalars couple preferentially to heavy fermions. The \ttbar associated production mode cross-section is proportional to the coupling of the light boson to top quarks, which can be modified in extended Higgs-sector models~\cite{Casolino:2015cza}, and provides information that complements searches relying on light resonance production from SM Higgs boson decays. Since mass-proportional couplings are largest for third-generation fermions, the decay channel to a $\tau^+\tau^-$ pair is a sensitive handle to explore such scenarios~\cite{Cacciapaglia:2017iws} and complements searches in other final states, such as the bottom--antibottom quark pair decay channel~\cite{HMBS-2024-46}.

The present study investigates a low-mass Higgs-like particle (denoted $a$) produced in association with a \ttbar pair, where the Higgs-like particle decays into a $\tauhad\tauhad$ (\dt) final state.\footnote{The \tauhad notation refers to the hadronic decay mode of the \tau-lepton, in which it decays into a neutrino, at least one charged hadron (most commonly pions), and possibly additional neutral hadrons. These hadrons comprise the \textit{visible} decay products of the \tauhad, denoted in this paper by the \textit{vis} subscript.}
The analysis focuses on the mass ($m_a$) region in the range of \SI{20}{\GeV} to \SI{85}{\GeV}, targeting a topology where the $a$-resonance receives a significant Lorentz-boost, rendering its decay products highly collimated and mandating the use of novel reconstruction techniques. The analysis sensitivity is enhanced by leveraging the boosted topology of the \dt final state, where both \tau-leptons from the Higgs-like particle decay are reconstructed as sub-jets within a single large-radius jet. This strategy facilitates the signal-background separation, and improves the reconstruction precision. The lower bound on the target $m_a$ range is driven by the increasingly challenging reconstruction, as average transverse momenta of individual \tau-lepton decay products decrease while the Lorentz-boost of the \dt system increases. The upper bound on the target $m_a$ range is constrained by
the reduction in the fraction of signal particles sufficiently boosted for their resultant \dt decays to be reconstructed as a single large-radius jet.

\begin{figure}
\centering
\includegraphics[width=0.4\textwidth]{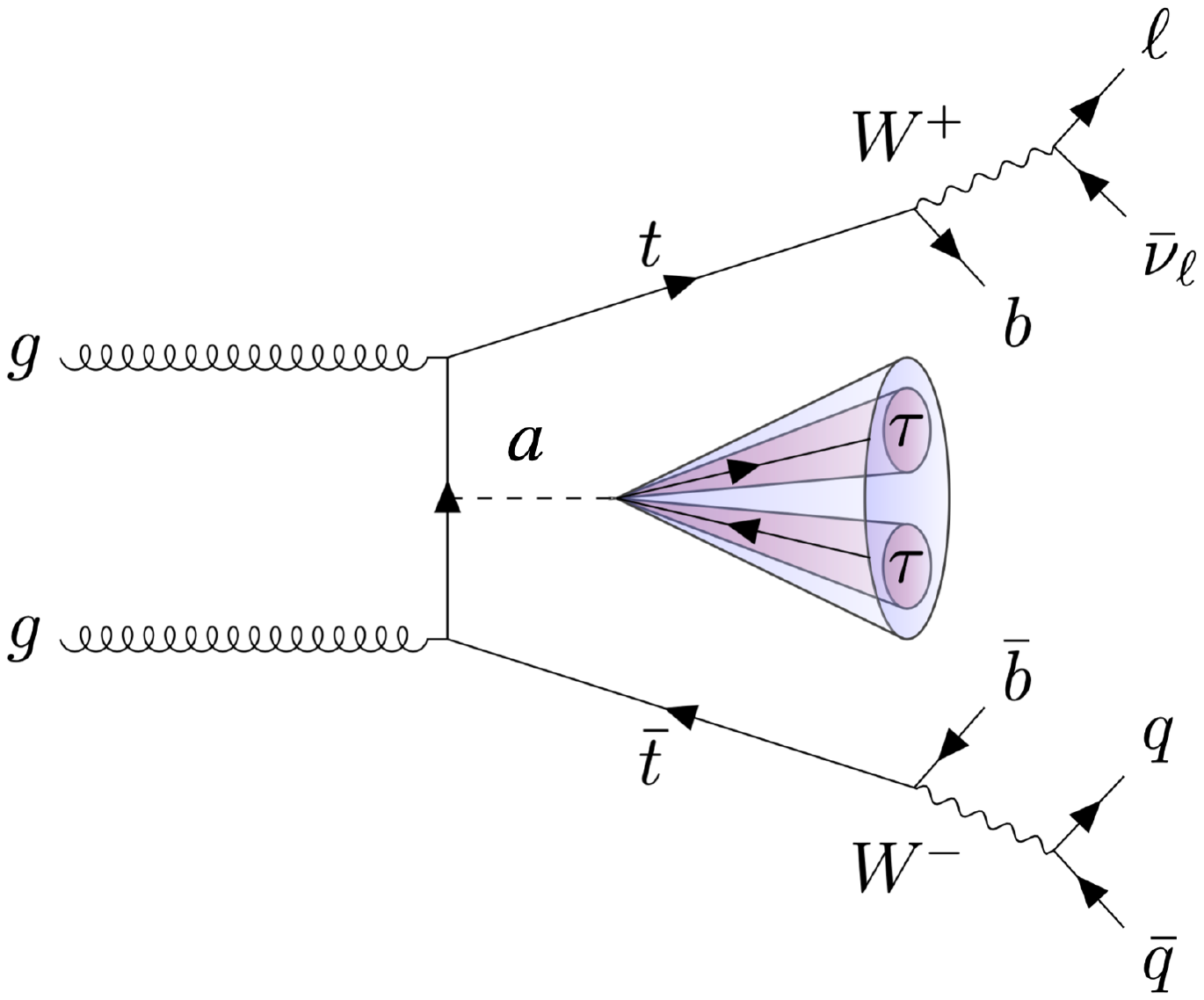}
\caption{Schematic representation of the signal process considered in this analysis. A low-mass resonance $a$ is produced in association with a \ttbar pair and decays into a fully hadronic \dt final state.}
\label{fig:introduction:tta_diagram}
\end{figure}

The semileptonic decay mode of the \ttbar{} system is targeted, in which one top-quark decays hadronically while the other undergoes a leptonic decay ($t\to W b \to l\nu_l b$) to an electron or a muon. The electron or muon from the top-quark decay provides an effective event trigger and contributes to background suppression. Advanced jet substructure techniques and boosted object tagging are employed to accurately reconstruct the fully hadronic \dt system in this $m_a$ region~\cite{TAUP-2020-03}, where only a very small irreducible SM contribution is expected. The diagram of the production and decay process is shown in Figure~\ref{fig:introduction:tta_diagram}. The dominant background arises from \ttbar{} production, particularly from events in which at least one jet is misreconstructed as a hadronically decaying $\tau$-lepton.

Previous searches for low-mass resonances in \dt final states have primarily focused on the $\tau_{\text{lep}}\tau_{\text{had}}$ decay channel~\cite{CMS-HIG-17-014,HDBS-2021-07,HMBS-2024-25}. The fully hadronic \dt final state has been considered in the context of a search for exotic Higgs boson decays into a pair of pseudoscalars in ATLAS~\cite{HMBS-2024-39}, while CMS has considered it in one of the channels in the context of a search for dilepton resonances in associated vector-boson or \ttbar production~\cite{CMS-EXO-21-018}. However, the $a\to\tauhadtauhad$ topology in \ttbar-associated production has largely remained unexplored by experiments in the highly boosted regime where resolved $\tauhad$ reconstruction becomes inefficient.
This analysis addresses this gap by specifically targeting the low $m_{\tau\tau}$ phase space region using such a boosted \dt topology, delivering the first ATLAS result in this final state. The results of this search provide further insights into potential extensions of the Higgs sector, contributing to the broader search for BSM physics at the LHC by complementing recent ATLAS searches for top-quark associated $a$ production in the $a\rightarrow\mu\mu$~\cite{HDBS-2020-12} and $a\rightarrow b\bar{b}$~\cite{HMBS-2024-46} decay channels. The dataset utilised in this study comprises \SI{140}{\per\femto\barn} of proton--proton (\pp) collision data collected at $\sqs=\SI{13}{\TeV}$ with the ATLAS detector during Run 2 (2015–2018).%

The paper is structured as follows. A brief discussion of the ATLAS detector and an overview of the Monte Carlo samples and data used in the analysis are presented in Sections~\ref{sec:detector} and~\ref{sec:data-mc}, respectively. Object reconstruction and the event selection are described in Section~\ref{sec:event-reco-sel}, with special focus on the custom reconstruction of boosted di-$\tau$ objects. The signal modelling and background estimates are discussed in Section~\ref{sec:background}, followed by a description of the statistical method and dominant systematic uncertainties in Section~\ref{sec:stats}. Finally, results are presented in Section~\ref{sec:results} and conclusions are drawn in Section~\ref{sec:conclusion}.

%

%

% End of text imported from the .//chapters/introduction.tex input file

%

%
\section{The ATLAS detector}
\label{sec:detector}

% The next lines are included from the .//chapters/atlas.tex input file
%
\newcommand{\AtlasCoordFootnote}{%
ATLAS uses a right-handed coordinate system with its origin at the nominal interaction point (IP)
in the centre of the detector and the \(z\)-axis along the beam pipe.
The \(x\)-axis points from the IP to the centre of the LHC ring,
and the \(y\)-axis points upwards.
Polar coordinates \((r,\phi)\) are used in the transverse plane,
\(\phi\) being the azimuthal angle around the \(z\)-axis.
The pseudorapidity is defined in terms of the polar angle \(\theta\) as \(\eta = -\ln \tan(\theta/2)\) and is equal to the rapidity
$ y = \frac{1}{2} \ln \left( \frac{E + p_z}{E - p_z} \right) $ in the relativistic limit.
Angular distance is measured in units of \(\Delta R \equiv \sqrt{(\Delta\eta)^{2} + (\Delta\phi)^{2}}\).}

The ATLAS detector~\cite{PERF-2007-01} at the LHC covers nearly the entire solid angle around the collision point.\footnote{\AtlasCoordFootnote} It consists of an inner tracking detector surrounded by a thin superconducting solenoid, electromagnetic and hadronic calorimeters, and a muon spectrometer incorporating three large superconducting air-core toroidal magnets.

The inner-detector system (ID) is immersed in a \qty{2}{\tesla} axial magnetic field and provides charged-particle tracking in the range \(|\eta| < 2.5\). The high-granularity silicon pixel detector covers the vertex region and typically provides four measurements per track,
the first hit generally being in the insertable B-layer (IBL) installed before Run~2~\cite{ATLAS-TDR-19,PIX-2018-001}. It is followed by the SemiConductor Tracker (SCT), which usually provides eight measurements per track. These silicon detectors are complemented by the transition radiation tracker (TRT), which enables radially extended track reconstruction up to \(|\eta| = 2.0\). The TRT also provides electron identification information based on the fraction of hits (typically 30 in total) above a higher energy-deposit threshold corresponding to transition radiation.

The calorimeter system covers the pseudorapidity range \(|\eta| < 4.9\). Within the region \(|\eta|< 3.2\), electromagnetic calorimetry is provided by barrel and endcap high-granularity lead/liquid-argon (LAr) calorimeters, with an additional thin LAr presampler covering \(|\eta| < 1.8\) to correct for energy loss in material upstream of the calorimeters. Hadronic calorimetry is provided by the steel/scintillator-tile calorimeter, segmented into three barrel structures within \(|\eta| < 1.7\), and two copper/LAr hadronic endcap calorimeters. The solid angle coverage is completed with forward copper/LAr and tungsten/LAr calorimeter modules optimised for electromagnetic and hadronic energy measurements respectively.

The muon spectrometer (MS) comprises separate trigger and high-precision tracking chambers measuring the deflection of muons in a magnetic field generated by the superconducting air-core toroidal magnets. The field integral of the toroids ranges between \num{2.0} and \qty{6.0}{\tesla\metre} across most of the detector. Three layers of precision chambers, each consisting of layers of monitored drift tubes, cover the region \(|\eta| < 2.7\), complemented by cathode-strip chambers in the forward region, where the background is the highest. The muon trigger system covers the range \(|\eta| < 2.4\) with resistive-plate chambers in the barrel, and thin-gap chambers in the endcap regions.

The luminosity is measured mainly by the LUCID--2~\cite{LUCID2} detector that records Cherenkov light produced in the quartz windows of photomultipliers located close to the beampipe.

Events are selected by the first-level trigger system implemented in custom hardware, followed by selections made by algorithms implemented in software in the high-level trigger~\cite{TRIG-2016-01}. The first-level trigger accepts events from the \qty{40}{\MHz} bunch crossings at a rate close to \qty{100}{\kHz}, which the high-level trigger further reduces in order to record complete events to disk at about \qty{1.25}{\kHz}.

A software suite~\cite{SOFT-2022-02} is used in data simulation, in the reconstruction and analysis of real and simulated data, in detector operations, and in the trigger and data acquisition systems of the experiment.
%

%

%

% End of text imported from the .//chapters/atlas.tex input file

%

%
\section{Data and simulated event samples}
\label{sec:data-mc}

% The next lines are included from the .//chapters/data_mc.tex input file
This search uses the $\sqrt{s} = \SI{13}{\TeV}$ \pp collision data collected by the ATLAS detector during the period from 2015 to 2018, referred to as the full Run 2 dataset. Only events recorded under stable beam conditions with all ATLAS subsystems fully operational  are considered~\cite{DAPR-2018-01}, %
corresponding to an integrated luminosity of $\lumir\pm\lumiruncertfb$~\cite{DAPR-2021-01}.

The dataset was collected using single-electron~\cite{TRIG-2018-05} and single-muon~\cite{TRIG-2018-01} triggers.
To maximise trigger efficiency, triggers with both isolated and non-isolated lepton requirements are used. As the luminosity increased during Run 2, the triggers were adjusted with progressively higher transverse momentum (\pT) thresholds to cope with increasing event rates. For electrons, the \pT threshold was set at $\SI{24}{\GeV}$ in 2015 for isolated triggers and increased to $\SI{26}{\GeV}$ during 2016--2018, using track- and calorimeter-based isolation criteria with variable and fixed cone sizes, respectively. Non-isolated electron triggers required higher \pT values, beginning at $\SI{60}{\GeV}$. Similarly, muon triggers required $\pT \geq \SI{20}{\GeV}$ in 2015, rising to $\SI{26}{\GeV}$ from 2016 onwards for isolated triggers with track-based variable-cone isolation, while non-isolated triggers consistently required $\pT \geq \SI{50}{\GeV}$. These adjustments ensured high trigger efficiency while controlling the data acquisition rate.

Monte Carlo (MC) simulated samples are used to model signal and background processes, and to derive related modelling uncertainties. The generated events are processed through a full simulation~\cite{SOFT-2010-01} of the ATLAS geometry and response based on \GEANT~\cite{Agostinelli:2002hh}.
The effect of multiple interactions in the same and neighbouring bunch crossings (\pileup) was modelled by overlaying the simulated hard-scattering event with inelastic \pp events generated with \PYTHIA[8.186]~\cite{Sjostrand:2007gs} using the \NNPDF[2.3lo] set of parton distribution functions (PDF)~\cite{Ball:2012cx} and the
A3 set of tuned parameters~\cite{ATL-PHYS-PUB-2016-017}.

Simulated events are weighted to reproduce the distribution of the average number of interactions per bunch crossing observed in data. All simulated events are reconstructed with the same  algorithms as those used for data.

Signal events are simulated using a Two-Higgs-Doublet Model (2HDM) framework, which serves as a simplified framework for generating events with light scalar resonances, independent of specific theoretical assumptions. Signal production is modelled at leading-order (LO) using \MGNLO[2.3.3]~\cite{Alwall:2014hca}, interfaced with \PYTHIA[8.212]~\cite{Sjostrand:2014zea} for parton showering, hadronisation and the \atautau decay, where the \tau-leptons are set to decay hadronically. The matrix element calculation was performed with the \NNPDF[2.3lo] PDF set. The renormalisation and factorisation scales were set to the \MADGRAPH default values, based  on a clustering of the event. The A14 tune~\cite{ATL-PHYS-PUB-2014-021} of \PYTHIA{8} was used with the \NNPDF[2.3lo]PDF set. The decays of bottom and charm hadrons were performed by \EVTGEN[1.2.0]~\cite{Lange:2001uf}. The mass of the pseudoscalar resonance $m_a$ is varied in steps of $\SI{5}{\GeV}$ for masses between $\SI{20}{\GeV}$ and $\SI{85}{\GeV}$.

The main background for this search is due to quark- or gluon-initiated jets that are misreconstructed as boosted \dt objects, originating predominantly from events that include the production of a top quark (mostly \ttbar and \SingleTop) or the production of vector bosons (\Vjets and \diboson). An irreducible background contribution originating from prompt boosted \dt objects is expected to arise from the \tttautau process. Additional subdominant background sources, grouped and denoted by ``Others'' in the figures and tables in this paper, include contributions from \ttZg, \ttW, \ttH, \ttt (3-top) and \tttt (4-top) processes.
The main \ttbar background process, filtered to exclude fully hadronic decays, was modelled using the \POWHEGBOX[v2]~\cite{Frixione:2007nw,Nason:2004rx,Frixione:2007vw,Alioli:2010xd} generator at next-to-leading-order (NLO) with the \NNPDF[3.0nlo]~\cite{Ball:2014uwa} PDF set and the \hdamp parameter\footnote{The \hdamp parameter is a resummation damping factor and one of the parameters that controls the matching of \POWHEG matrix elements to the parton shower and thus effectively regulates the high-\pT radiation against which the \ttbar system recoils.} set to 1.5\mtop~\cite{ATL-PHYS-PUB-2016-020}. The events were interfaced with \PYTHIA[8.230] to model the parton shower, hadronisation, and underlying event, with parameters set according to the A14 tune and using the \NNPDF[2.3lo] set of PDFs. The \NNPDF[3.0lo] replicas were used to evaluate the PDF uncertainties for the nominal PDF\@. In addition, the central value of this PDF was compared with the central values of the \CT[14nnlo]~\cite{Dulat:2015mca} and \MMHT[nnlo]~\cite{Harland-Lang:2014zoa} PDF sets. To assess the uncertainty in the matching of NLO matrix elements to the parton shower, the \POWHEG sample was compared with a sample of events generated with \MGNLO[2.6.0] interfaced with \PYTHIA[8.230]. The \MGNLO calculation used the \NNPDF[3.0nlo] set of PDFs and \PYTHIA[8] used the A14 set of tuned parameters and the \NNPDF[2.3lo] set of PDFs. The impact of using a different parton shower and hadronisation model was evaluated by comparing the nominal \ttbar sample with another event sample produced with the \POWHEGBOX[v2]generator using the \NNPDF[3.0nlo] parton distribution function. Events in the latter sample were interfaced with \HERWIG[7.13]~\cite{Bahr:2008pv,Bellm:2015jjp}, using the \HERWIG[7.1] default set of tuned parameters~\cite{Bellm:2015jjp,Bellm:2017jjp}. In all \ttbar samples, the decays of bottom and charm hadrons were simulated using the \EVTGEN[1.6.0] programme.

Single-top backgrounds were simulated at NLO using \POWPYeight for the $t$-, $s$-, and $Wt$-channels, while an additional $t$-channel sample was generated with \MGNLO interfaced with \PYTHIA{8}. In all cases, \PYTHIA was used with the \NNPDF[2.3lo] PDF set with the A14 tune for parton showering. The \ttt and \tttt processes were generated at LO using \MGNLO and \PYTHIA{8}, with the \NNPDF[2.3lo] PDF applied in both the matrix element generation and parton showering stages.

The associated production of a vector boson (\ttW and \ttZ) was simulated with \MGNLO and \PYTHIA{8} at NLO, using the \NNPDF[2.3lo] PDF\@. In particular, for the \tttautau background process, alternative samples simulated with \SHERPA[2.2.1]~\cite{Bothmann:2019yzt} were used for comparison to the nominal samples in order to estimate hard-scatter process uncertainties. The \ttH process was modelled with \POWPYeight at NLO, using the \NNPDF[3.0nlo] PDF and the A14 tune. The \Zjets and \Wjets backgrounds were simulated with \SHERPA 2.2.1, employing the \NNPDF[3.0nnlo] PDFs for both matrix element generation and parton showering. Events with up to two jets were generated at NLO, while events with three or more jets were generated at LO\@. The samples were further sliced by $\max(\HT, \pT^V)$ and jet-flavour filters to enhance phase-space coverage, where $\HT$ is defined as the scalar \pT sum of all jets in the event and $\pT^V$ is the transverse momentum of the vector boson.
\Diboson  events in the \semilep and \fullylep decay channels were simulated with \POWHEG interfaced with \PYTHIA{8}, with the \NNPDF[3.0nlo] PDF set, consistent with the configuration used for \ttbar simulation.

% End of text imported from the .//chapters/data_mc.tex input file

%

%
\section{Event reconstruction and selection}
\label{sec:event-reco-sel}
The analysis relies on a common set of reconstructed physics objects together with a dedicated algorithm for tagging boosted \dt candidates, and applies a selection focused on the associated semileptonic \ttbar topology. Events are preselected by requiring the presence of electrons or muons, small-radius jets, $b$-tagged jets, missing transverse momentum and boosted \dt candidates. Based on these objects, a signal region targeting a generic \tta production is defined, together with control and validation regions used to constrain and verify the background estimates.
\subsection{Event reconstruction}
\label{sec:event-reco}

% The next lines are included from the .//chapters/event_reconstruction.tex input file
Primary vertices are formed from sets of two or more tracks, each with transverse momentum $\pT > \SI{0.5}{\GeV}$, that are consistent with having originated at the same three-dimensional space point within the region of the colliding proton beams~\cite{ATL-PHYS-PUB-2015-026}. When more than one such primary vertex is found the vertex with the largest scalar sum of the squared transverse momenta of the associated tracks is chosen.

Electrons are reconstructed by matching energy clusters in the electromagnetic (EM) calorimeter to reconstructed tracks from the ID~\cite{EGAM-2018-01}. Selected electrons must have a transverse momentum $\pT > \SI{27}{\GeV}$ to satisfy trigger thresholds, and a pseudorapidity $|\eta| < 2.47$. To ensure high-quality electron reconstruction, candidates in the barrel--endcap transition region ($1.37 < |\eta| < 1.52$) are excluded. Selected electrons must satisfy \textit{Tight} identification criteria~\cite{PERF-2017-01}, with a longitudinal impact parameter $|z_0 \sin \theta| < \SI{0.5}{\mm}$ and transverse impact parameter significance $|d_0|/\sigma_{d_0} < 5$, both relative to the selected primary vertex. \textit{Tight} isolation criteria, involving tracking and calorimetry information with \pT-dependent and fixed cone radii, respectively, are also applied~\cite{EGAM-2021-02}.

Muons are reconstructed by combining information from tracks in the MS and the ID~\cite{MUON-2018-03}. Selected muons are required to have $\pT > \SI{27}{\GeV}$ to satisfy trigger thresholds, and a pseudorapidity $|\eta| < 2.4$. Muons must satisfy the \textit{Medium} identification criteria~\cite{MUON-2018-03}, with longitudinal impact parameter and transverse impact parameter significance relative to the selected primary vertex satisfying $|z_0 \sin \theta| < \SI{0.5}{\mm}$ and $|d_0|/\sigma_{d_0} < 3$, respectively. Isolation is enforced using tracking and calorimeter information, with \pT-dependent and fixed cone radii, respectively~\cite{MUON-2018-03}.

Jets are reconstructed with the \antikt algorithm~\cite{Cacciari:2008gp}, as implemented in \texttt{FastJet}~\cite{Fastjet}, using a jet radius parameter of $R = 0.4$. Input objects for this algorithm are particle-flow objects~\cite{PERF-2015-09}, combining measurements from the ATLAS inner detector and calorimeters~\cite{PERF-2014-07}. This approach enhances jet energy resolution and improves reconstruction efficiency, particularly for low-\pT jets. Selected jets are required to have $\pT > \SI{20}{\GeV}$ and $|\eta| < 2.8$. To further reduce the pile-up contamination, an additional selection using the jet-vertex-tagger~\cite{PERF-2014-03} (JVT) discriminating variable is employed, applying a requirement of $\text{JVT} > 0.5$ on jets that satisfy $\pT < \SI{60}{\GeV}$ and $|\eta| < 2.4$.

A $b$-tagging algorithm, known as ``DL1r''~\cite{FTAG-2019-07}, is used to identify jets containing $b$-hadrons. The DL1r algorithm uses a deep neural network combining information from several specialized low-level taggers using information from the interaction point, the impact parameters of tracks, and secondary and tertiary decay vertices to determine the probability for a jet to be $b$-, $c$-, or light-flavoured. The search employs the DL1r tagger at the 77\% efficiency working point, corresponding to the approximate average $b$-tagging efficiency in an inclusive \ttbar MC sample. %

Corrections are applied to the simulation to bring the efficiency of lepton reconstruction and identification in agreement with the data and to correct the energy (momentum) scale and resolution of electrons~\cite{EGAM-2021-02} (muons~\cite{MUON-2022-01}). Corrections are also applied to the simulation to bring the jet energy scale and jet energy resolution into agreement with the corresponding measurements in data~\cite{JETM-2018-05}, and measurements of the efficiency of the $b$-tagging algorithm are used to adjust the simulation to match the performance measured in data~\cite{FTAG-2018-01,FTAG-2019-02,FTAG-2020-08}.

Boosted \dt objects are reconstructed via a dedicated algorithm~\cite{TAUP-2020-03} from the substructure of a large-radius jet. The large-radius jet is reconstructed from particle-flow objects using the \antikt algorithm with a radius parameter of $R = 1.0$, and is required to satisfy $\pT > \SI{50}{\GeV}$ and $|\eta| < 2.5$. The large-radius jet constituents are subsequently reclustered into sub-jets with $\pT > \SI{10}{\GeV}$ using the \antikt algorithm with a radius parameter of $R = 0.2$, requiring at least two such sub-jets. The sub-jets are ordered in \pT, and the two highest-\pT sub-jets are referred to as the \textit{leading} and \textit{sub-leading} sub-jets, respectively. The leading and sub-leading sub-jets are assumed to contain the individual visible \tauhad signatures, thus the reconstructed \dt system's four-momentum is calculated as the vector sum of their individual four-momenta. The leading and sub-leading sub-jets are each required to have one or three associated tracks and their charge product is required to satisfy $\Qtt \equiv q_{\text{lead}} \times q_{\text{sub-lead}} = \pm 1$, where $q_{\text{lead}}$ ($q_{\text{sub-lead}}$) is defined as the sum of associated track charges of the leading (sub-leading) sub-jet. The leading and sub-leading sub-jets individually receive an energy calibration (Tau Energy Scale). The \pT of the sub-jet is first corrected to mitigate \pileup contributions. The \pileup-corrected sub-jet \pT is then adjusted to account for the response of the detectors. The calibrated \dt system's \pT is required to be greater than \SI{25}{\GeV} and its invariant mass (\mtt) is required to be between \SI{10}{\GeV} and \SI{60}{\GeV}. The \pT requirement is placed in order to suppress boosted \dt objects seeded by pile-up-dominated large-radius jets. The lower bound on \mtt is set below the minimal \ma considered in the analysis, to account for the missing momentum contributions from final-state neutrinos originating in the \tau-lepton decays.
The upper bound is also set below the maximal \ma considered, to both account for the missing momenta and to suppress the \ttZtautau background process contribution.

\begin{figure}[ht]
\centering
\includegraphics[width=0.49\linewidth]{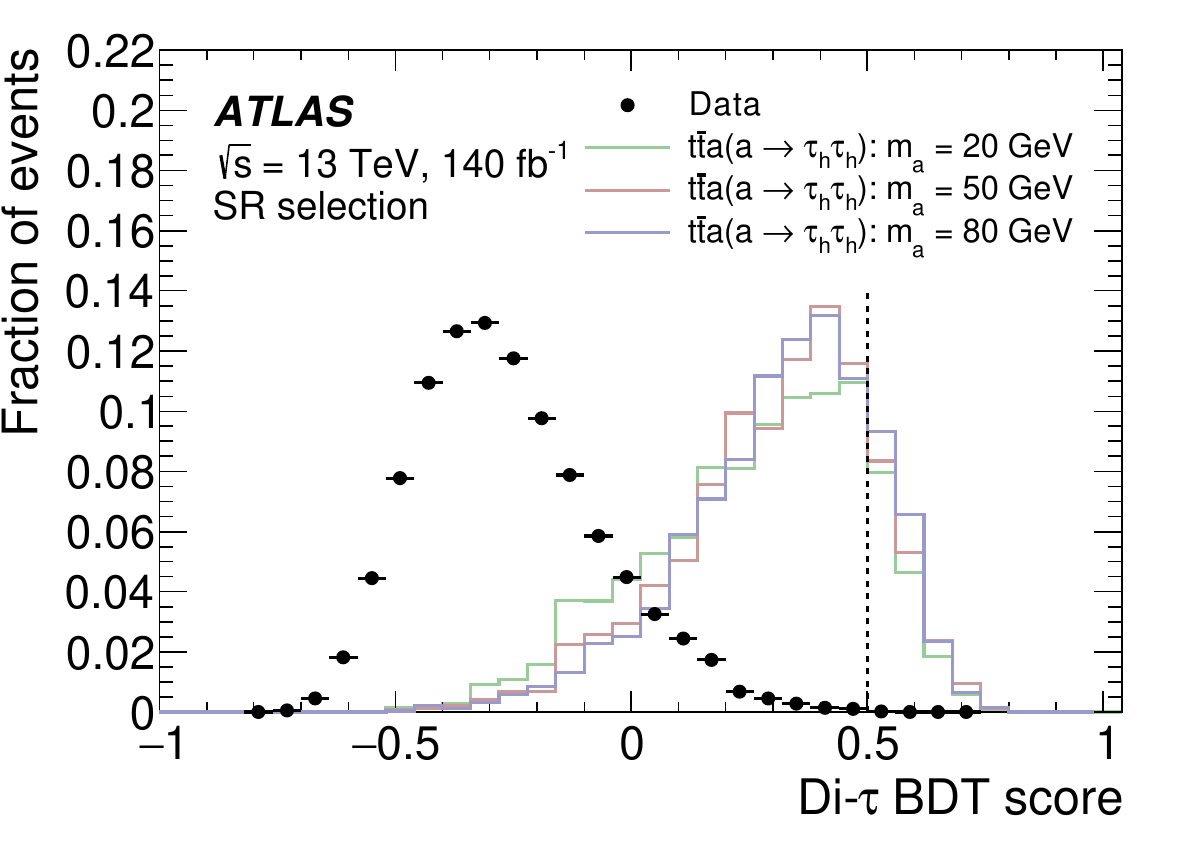}
\caption{Boosted \dt BDT score distribution for data and signal samples with $m_a$ values of \SI{20}{\GeV}, \SI{50}{\GeV}, and \SI{80}{\GeV}, for events satisfying the signal region selection criteria outlined in \cref{sec:event-selection}. Only signal events with prompt boosted \dts are included. The dotted line denotes a BDT score of 0.5, the minimum score required for boosted \dt objects included in the search signal region, corresponding to a signal efficiency of 15\%, 17\% and 19\% for each signal sample respectively, and a background rejection of 99.96\%. The distributions are normalised to unit area.}
\label{fig:event-reco:ditau:bdt}
\end{figure}

A dedicated identification algorithm is applied to distinguish prompt boosted \dt objects from misreconstructed (fake) ones originating from quark- or gluon-initiated jets, using a Boosted Decision Tree (BDT) based on a set of track- and calorimeter-based variables, trained using \SI{20}{\GeV} and \SI{60}{\GeV} \tta events as signal, against \ttbar background in a fully hadronic decay mode~\cite{TAUP-2020-03}. The boosted \dt BDT score distributions for both data and simulated signal samples corresponding to \ma values of \SI{20}{\GeV}, \SI{50}{\GeV} and \SI{80}{\GeV} are shown in \cref{fig:event-reco:ditau:bdt}, including events with the signal region selection (with the \dt BDT cut removed) described in \cref{sec:event-selection}. The performance of the reconstruction and identification algorithms is estimated in a tag-and-probe calibration analysis using boosted \dt objects from SM $Z(\to \tauhad\tauhad) + \gamma$ events~\cite{TAUP-2020-03}. Boosted \dt objects included in this search are required to satisfy an identification BDT score greater than 0.2, corresponding to an estimated signal efficiency of approximately 77\% at a background rejection factor of approximately 55.

A two-step overlap-removal procedure is applied to resolve ambiguities where multiple preselected physical objects are reconstructed from the same detector signature. The first step sequentially resolves ambiguities between electrons, muons and jets, based on the angular separation and properties of the reconstructed objects. This step is summarised in~\cref{tab:event-reconstruction:olr}.
In the second step, boosted \dt objects are discarded if overlapping with an electron or muon, and jets are discarded if overlapping with a boosted \dt object. The second step considers objects within $\DeltaR < 1$ of each other.

\begin{table}[ht]
\centering
\caption{Summary of the overlap removal procedure applied to resolve ambiguities between selected reconstructed electrons, muons and jets. The procedure is sequential, with each row of the table denoting the criteria applied to objects that are retained after all preceding steps.}
\label{tab:event-reconstruction:olr}
\begin{tabular}{lll}
\toprule
Remove & Retain & Criteria \\
\midrule
Electron (\pTX[][1]) & Electron (\pTX[][2]) & Shared track, \(\pTX[][1] < \pTX[][2]\) \\
Muon     & Electron & Shared inner detector track and the muon is calorimeter-tagged \\
Electron & Muon     & Shared inner detector track \\
Jet      & Electron & \(\Delta R < 0.2\) \\
Electron & Jet      & \(\Delta R < 0.4\)\\
Jet      & Muon     & \(\Delta R < 0.2\) and the jet has at most two associated tracks \\
Muon     & Jet      & \(\Delta R < \min(0.4, 0.04 + \qty{10}{\GeV}/\pT(\mu))\) \\
\bottomrule
\end{tabular}
\end{table}

The missing transverse momentum (with magnitude \MET) is reconstructed as the negative vector sum of the transverse momenta of all the reconstructed and calibrated objects in the event. The calculation includes a soft term that accounts for all tracks associated with the primary vertex but not matched to any reconstructed object~\cite{JETM-2020-03}. This calculation is performed between the first and second overlap-removal steps outlined previously; boosted \dt objects are not explicitly included to avoid double-counting their contributions. Events with more than one boosted \dt object satisfying the object selection are very rare, comprising less than one percent of the expected yield at this stage. If overlapping with an electron or muon, the boosted \dt object would have been discarded and not included in the missing momentum calculation. Conversely, since an overlap between a boosted \dt object and jets is extremely common, these overlapping jets (which would have been discarded) serve as indicators for the boosted \dt object's contribution to the calculation.

% End of text imported from the .//chapters/event_reconstruction.tex input file

%

%
\subsection{Event selection and categorization}
\label{sec:event-selection}

% The next lines are included from the .//chapters/event_selection.tex input file
The analysis is designed to compare the observed event yields with the SM predictions in a signal region (SR) optimised to enhance the presence of a generic \tta signal with low $m_a$. To minimise model dependence, the SR selection is defined by the expected physical characteristics and boosted topology of the target final state signature. Signal events are required to contain exactly one lepton (electron or muon), geometrically matched to a corresponding trigger-level object of the same flavour, and exactly one boosted \dt. The semileptonic decay of the associated \ttbar\ pair is further targeted by requiring at least three jets, with at least one jet identified as $b$-tagged.

Opposite charges are required in the leading and sub-leading sub-jets of the boosted \dt object ($\Qtt = -1$), corresponding to a neutral resonance. Furthermore, a minimum identification BDT score of 0.5 is required on the reconstructed boosted \dt object to minimise the presence of fakes from the SM background, corresponding to an estimated signal efficiency between 15\% and 20\% for all masses included in the analysis, at a background rejection factor of approximately 3000. Finally, to account for the neutrinos from the \dt and top-quark decays, the \MET is required to be greater than \SI{30}{\GeV}, and to satisfy $\dPhidtMet < 2$ to be preferentially aligned with the boosted \dt object. Selection criteria are summarised in \cref{tab:event-selection:regions}.

Preselected events failing to meet the SR selection criteria are retained and assigned to one of several control and validation regions for background modelling studies, validation of the analysis strategy, and estimation of dedicated uncertainties, as outlined in \cref{sec:background,sec:stats}. The validation regions are labelled according to the boosted \dt object BDT score range (\textit{L}, ``\textit{loose}'', for BDT$ \in [0.2, 0.35)$; \textit{M}, ``\textit{medium}'', for BDT$ \in [0.35, 0.5)$; \textit{T}, ``\textit{tight}'', for BDT$ \in [0.5, 1.0)$) and according to the sub-jet charge product sign (opposite sign (\textit{OS}) for $\Qtt = -1$, and same sign (\textit{SS}) for $\Qtt = +1$). Regions enriched in non-prompt leptons (fake electrons or muons) are denoted with the F$\ell$ suffix.

\begin{table}[ht]
\centering
\caption{Selection criteria for the control (\CRttbar for the \ttbar background normalisation, and \CRFL for the fake-lepton background estimation), validation, and signal regions. The criteria that make the regions orthogonal to the SR are shown in boldface. The validation regions are labelled according to the boosted \dt object BDT score range (\textit{L} for BDT$ \in [0.2, 0.35)$, \textit{M} for BDT$ \in [0.35, 0.5)$, and \textit{T} for BDT$ \in [0.5, 1.0)$) and sub-jet charge product sign (\textit{OS} for $\Qtt = -1$, and \textit{SS} for $\Qtt = +1$). Regions enriched in non-prompt leptons (fake electrons or muons) are denoted with the F$\ell$ suffix.}
\label{tab:event-selection:regions}
\begin{tabular}{l cccccc}
\toprule
& \Njet     & \Nbjet    & \MET (\si{\GeV})  & \dPhidtMet    & \Qtt      & \Dt BDT Score          \\
\midrule
\CRttbar    & $\geq 3$  & $\geq 1$  & $> 30$            & --             & $\mathbf{+1}$ & $\mathbf{[0.2~  , 0.35)}$   \\
\CRFL       & $\geq 1$  & $\geq 1$  & $ \mathbf{\leq 30}$       & --             & $\mathbf{+1}$ &      $[0.2~  , 1   )$   \\
\midrule
\VRFL       & $\geq 1$  & $\geq 1$  & $\mathbf{ \leq 30}$       & --             &      $-1$ &      $[0.2~  , 1   )$   \\
\VRLOS      & $\geq 3$  & $\geq 1$  & $> 30$            & --             &      $-1$ & $\mathbf{[0.2~  , 0.35)}$   \\
\VRMSS      & $\geq 3$  & $\geq 1$  & $> 30$            & --             & $\mathbf{+1}$ & $\mathbf{[0.35~ , 0.5 )}$   \\
\VRMOS      & $\geq 3$  & $\geq 1$  & $> 30$            & --             &      $-1$ & $\mathbf{[0.35~ , 0.5 )}$   \\
\VRTSS      & $\geq 3$  & $\geq 1$  & $> 30$            & $< 2$         & $\mathbf{+1}$ &      $[0.5~  , 1   )$   \\
\midrule
\SR         & $\geq 3$  & $\geq 1$  & $> 30$            & $< 2$         &      $-1$ &      $[0.5~  , 1   )$   \\
\bottomrule
\end{tabular}
\end{table}

% End of text imported from the .//chapters/event_selection.tex input file

%

%
\section{Background estimation}
\label{sec:background}

% The next lines are included from the .//chapters/background_estimation.tex input file
The application of the signal-region selection criteria results in a SM background contribution dominated by events wherein jets mimic the signature of either the \dt, the lepton, or both. In particular, the scarcity of natural boosted \dt resonances in the SM implies that nearly all ($\gtrsim 99\%$) background events in the total analysis phase space (including the signal, control, and validation regions) contain a fake boosted \dt object, reconstructed from one or two nearby quark- or gluon-initiated jets.

Approximately 65\% of all SR background events with prompt leptons and fake \dts are expected to correspond to SM \ttbar production, with secondary contributions from \Vjets, \SingleTop and \diboson processes. The contribution from events with fake leptons is minimised by the strict identification and isolation requirements applied to the electrons and muons included in the analysis. In particular, the fraction of events with a fake lepton and a prompt boosted \dt -- originating predominantly from \Ztautau decays in SM high-\pT \Zjets events -- accounts for less than 0.5\% of the total yield in the SR\@. Events with both a fake \dt and a fake lepton originate most commonly from pure QCD multijet processes, which remain present in the SR despite the strict object quality criteria as a result of the very large production cross-sections, comprising approximately 8\% of the expected background in the SR\@. Finally, an 8\% SM contribution from the \tttautau process is expected in the SR, mostly from events with both prompt boosted \dt objects and prompt leptons. The boosted \dt purity in \tttautau events exceeds 90\%, and these events closely resemble the signal process' topology.

\subsection{\ttbar modelling}

With a very large cross-section and a similar final state to the signal process, the \ttbar background has the highest predicted yield of all SM processes in the SR\@. The background's kinematics are estimated through MC simulations, and normalised to data in a dedicated control region (\CRttbar). The \CRttbar selection criteria are included in \cref{tab:event-selection:regions}. Due to the partially shared final state topology, \CRttbar is defined with a similar selection to the SR, but requiring a boosted \dt object with same-sign ($\Qtt = +1$) sub-jets in an orthogonal BDT score interval ($0.2 < \text{BDT} < 0.35$) -- minimising any signal contamination to below the level of $0.01\%$. The \dPhidtMet selection applied in the signal region is removed, as the boosted \dt object is not expected to originate from prompt \tau-leptons and, therefore, is no longer a source of neutrinos. The \MET selection, however, is retained in order to account for neutrinos from top-quark decays while suppressing other QCD contributions, resulting in a \ttbar event purity of approximately 77\%. The remaining secondary SM backgrounds with dominant prompt-lepton contributions are modelled entirely from MC simulations.

\subsection{Background model validation}

The background modelling is further validated in a set of validation regions, denoted with the \textit{VR} prefix, obtained by varying the boosted \dt object selection criteria, also defined in \cref{tab:event-selection:regions}. These regions (VRLOS, VRMSS, VRMOS, and VRTSS) are defined based on the \CRttbar selection criteria, and enable tests of the extrapolation of the background estimates across the different combinations of \Qtt values and BDT score regions present in the intermediate phase space between the \CRttbar and the SR\@. The signal-like VRTSS further includes the \dPhidtMet selection also implemented in the search SR, but is kept orthogonal by requiring $\Qtt = 1$.

\begin{figure}[ht]
\centering
\includegraphics[width=0.5\linewidth]{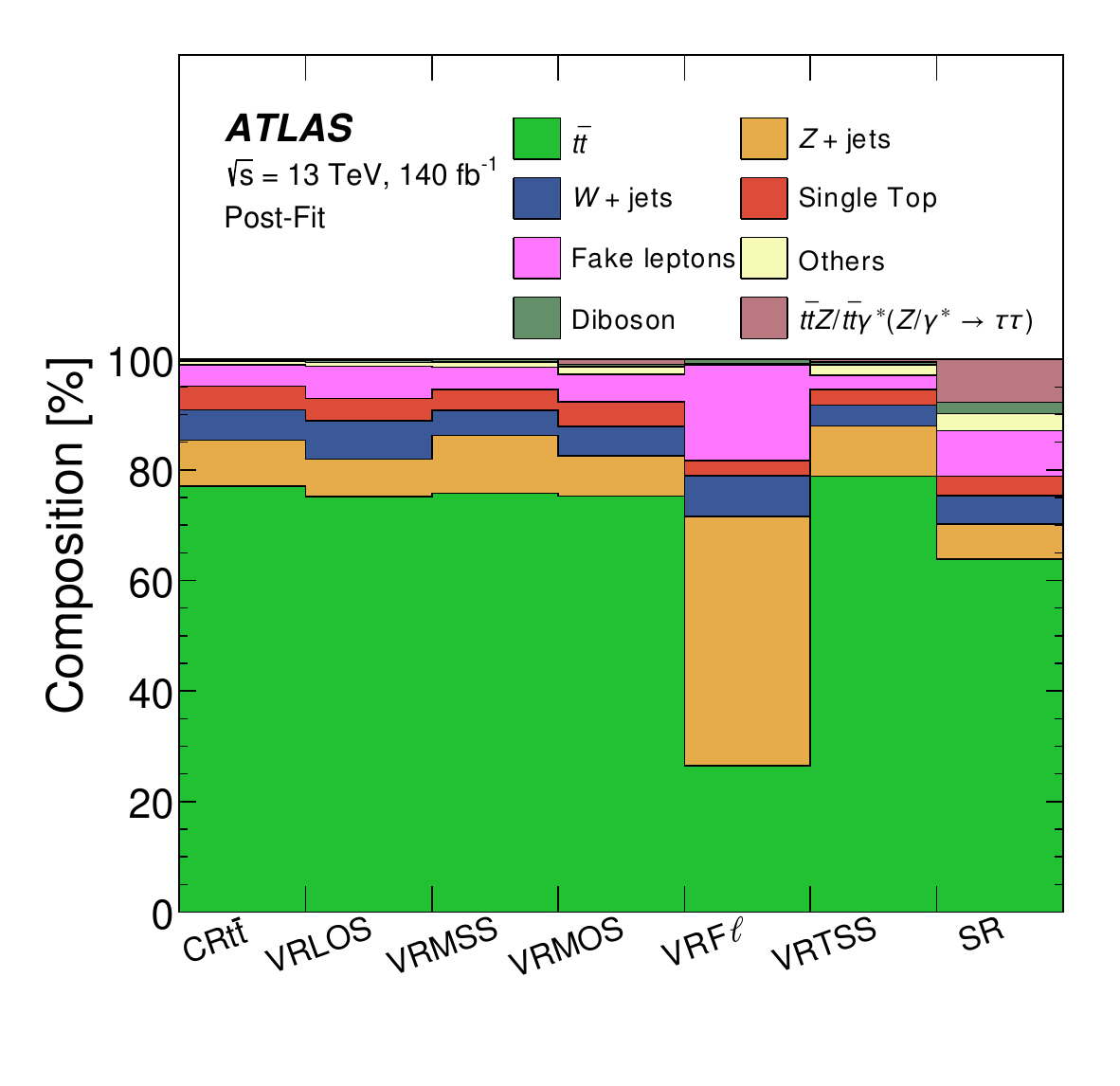}
\caption{Background composition of the analysis control, validation and signal regions, after a background-only fit. Minor background processes, such as $t\bar{t}(Z \to \ell^+\ell^-/ q\bar{q})$, \ttW, \ttH, 3 top and 4 top productions are considered under `Others'.}
\label{fig:background:composition}
\end{figure}

Although the \tttautau component with prompt boosted \dts is uniquely present in the SR, the overall background composition in the \CRttbar and the non-fake-enriched validation regions is broadly similar to that of the SR, apart from a small (\(\sim\)5\%) contribution from \diboson and other minor processes, as illustrated in \cref{fig:background:composition}.

\subsection{Misidentified lepton background estimate}\label{sec:stats:fakelep}

A dedicated data-driven estimation of events with fake leptons and fake boosted \dts is performed using the Matrix Method~\cite{EGAM-2019-01}. The lepton misidentification efficiencies, parameterised by the lepton's \pT and $|\eta|$, are measured in a multijet-enriched auxiliary control region (\CRFL) defined by requiring $Q_{\tau\tau}=1$ and inverting the \MET requirement used in the signal region to require $\MET\leq \SI{30}{\GeV}$. The lowest-threshold single-lepton triggers used in the analysis impose a high baseline isolation, severely restricting the number of events with isolation-inverted leptons (failing to meet the \textit{Tight} selection described in \cref{sec:event-reco}). Therefore, a looser boosted \dt object selection is applied, including all same-signed candidates in a wide $0.2 < \text{BDT} < 1.0$ discriminant score range. Prompt-identification efficiencies, also parameterised in bins of the lepton's \pT and $|\eta|$, are measured in the region defined by inverting this \MET requirement. Furthermore, to avoid negative estimates of fake-lepton event yields arising as an artifact of statistical limitations, the likelihood-based implementation of the Matrix Method is used, which guarantees non-negative event yield estimates in all regions~\cite{EGAM-2019-01} by construction.

\begin{figure}
\centering
\subfloat[]{\label{fig:background:VRFL:el}\includegraphics[width=0.49\linewidth]{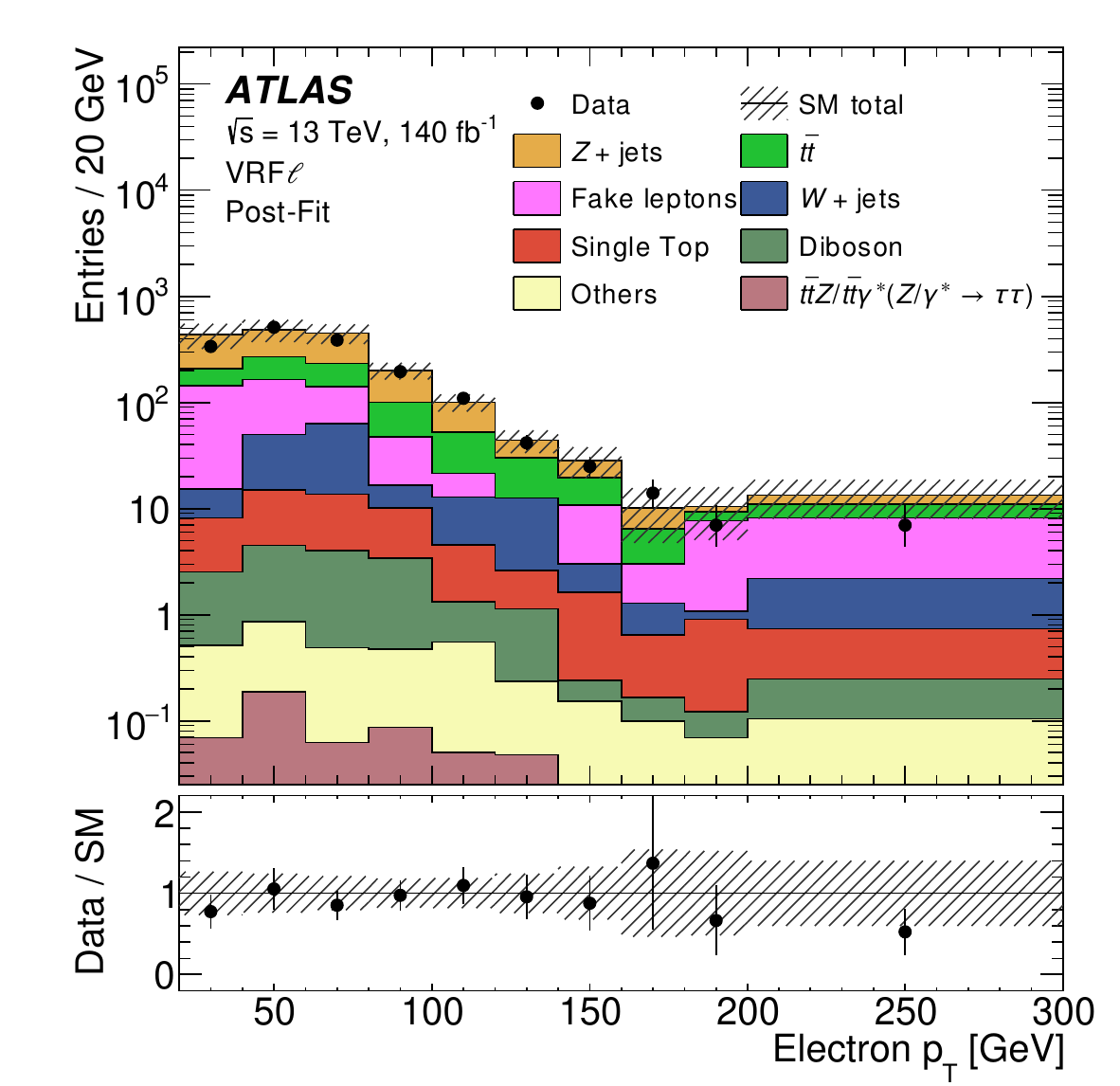}}
\subfloat[]{\label{fig:background:VRFL:mu}\includegraphics[width=0.49\linewidth]{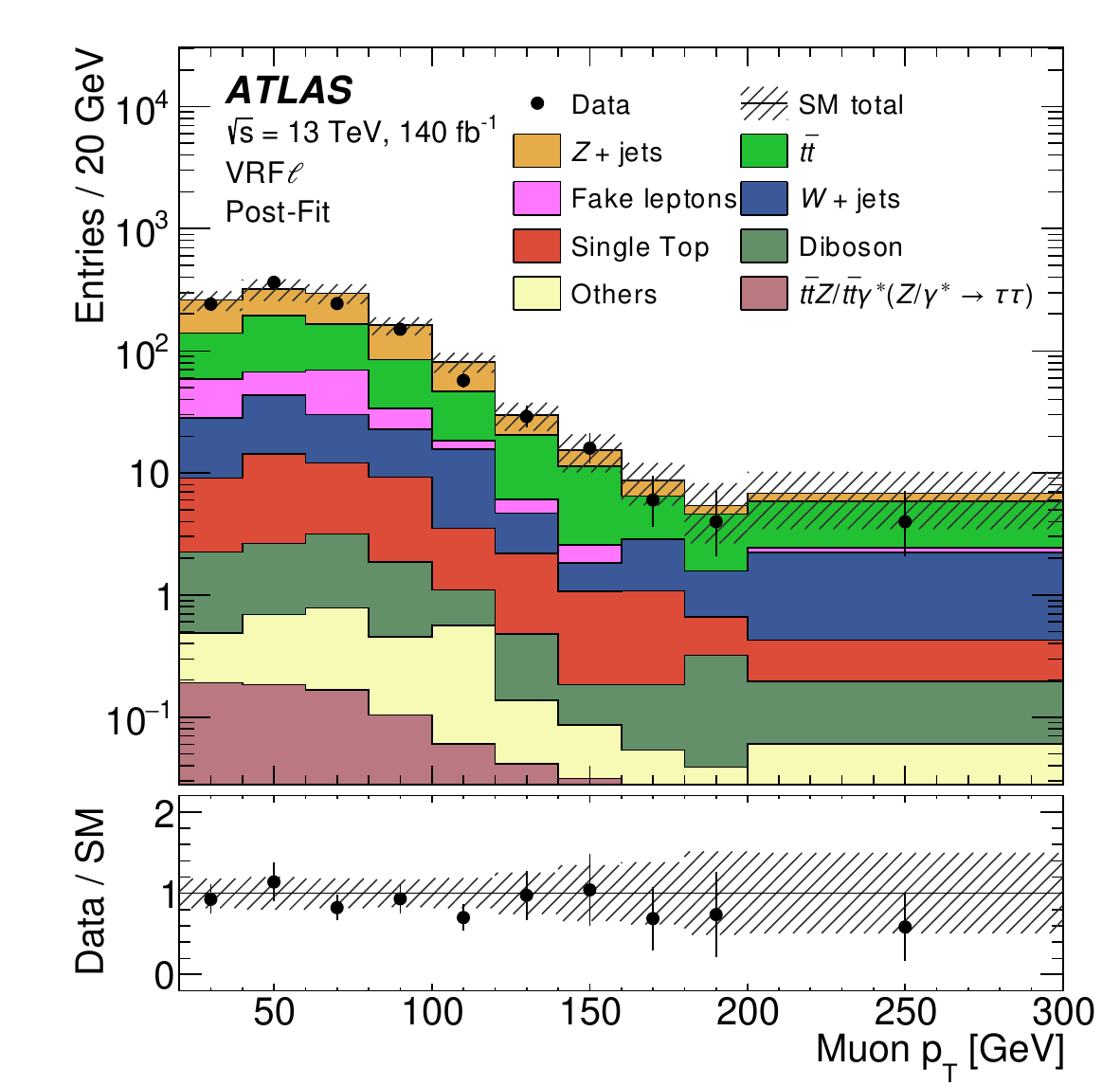}}
\caption{Observed and expected distributions of the electron \protect\subref{fig:background:VRFL:el} and muon \protect\subref{fig:background:VRFL:mu} \pT in the multijet-enriched fake-lepton validation region, after a background-only fit in the \CRttbar. Statistical and systematic uncertainties are included in the background estimate uncertainty bands. Overflow events are included in the last bins. Minor background processes, such as $t\bar{t}(Z/\gamma^* \to \ell^+\ell^-/q\bar{q})$, \ttW, \ttH, 3 top and 4 top productions are considered under `Others'. The lower panels show the ratio of the data to the total SM background prediction.}
\label{fig:background:VRFL}
\end{figure}

The resulting fake-lepton background estimate is validated in an orthogonal fake-enriched region (\VRFL) by inverting the \Qtt selection, shown in \cref{fig:background:VRFL}. The main background contribution in this region, comprising approximately 45\% of the total predicted yield, originates from \Zjets events with prompt leptons and fake boosted \dts. Since both the \CRFL and \VRFL are completely dominated by fake boosted \dt events, and fake \dt objects have no charge preference, flipping the \Qtt sign does not appreciably change the predicted SM composition between the two regions. After including the contribution from the Matrix Method estimate, the background estimate is consistent with observed data in the \VRFL across the lepton \pT spectrum for both electrons and muons. Figures~\ref{fig:background:VRFL:el} and~\ref{fig:background:VRFL:mu} show the fractional contribution of fake-lepton events to the total background estimate in the electron and muon channels, respectively. A smaller fake-lepton contribution is observed in the muon channel.
Quantitatively, the modelling of the background relative to the data in the \VRFL results in a deviation between total yields amounting to approximately 8\%. While the trigger-induced isolation bias does limit the fake-lepton purity in the \VRFL,
it does not constitute a problem due to the low expected fake-lepton contribution in the analysis SR and the dedicated uncertainty attributed to the fake-lepton estimate (detailed in Section~\ref{sec:stats}).

% End of text imported from the .//chapters/background_estimation.tex input file

%

%
\section{Statistical analysis}
\label{sec:stats}

% The next lines are included from the .//chapters/statistical_analysis.tex input file
%

Likelihood fits~\cite{Baak:2014wma} are performed under three configurations: i) a background-only hypothesis, where the fit is performed in the \CRttbar region only and is used to estimate the expected background in the SR and VRs; ii) a model-dependent (exclusion) signal-plus-background hypothesis, where the fit is performed simultaneously in the \CRttbar and SR regions; and iii) a model-independent (discovery) signal-plus-background hypothesis, where a single-bin fit is performed simultaneously in the \CRttbar and SR, while injecting into the SR a single signal event without associated uncertainties. In the signal-plus-background configurations, a signal strength parameter $\mu_{\text{sig}}$ is introduced, and a free normalisation factor for the \ttbar background, $\mu_{\ttbar}$, is floated simultaneously in the \CRttbar and SR regions. In the background-only configuration, $\mu_{\ttbar}$ is determined exclusively in the \CRttbar region and extrapolated to the SR and VRs.

In the model-dependent interpretation, additional discriminating power is obtained by subdividing the SR into three \mtt bins, improving the sensitivity to low-mass resonance signals while maintaining sufficient statistical precision in the background prediction. The three \mtt intervals are defined as $\SI{10}{\GeV} \leq \mtt < \SI{20}{\GeV}$, $\SI{20}{\GeV} \leq \mtt < \SI{30}{\GeV}$, and $\SI{30}{\GeV} \leq \mtt < \SI{60}{\GeV}$.

A test statistic based on the profile likelihood ratio~\cite{Cowan:2010js} is used to test hypothesized values of the signal strength, $\mu_\signal$, for each of the simulated mass points in the SR\@. In the absence of signal, the expected and observed 95\% confidence level (CL) exclusion limits are evaluated based on the $\text{CL}_s$ method~\cite{Read:2002hq}. Upper limits on the fitted $\mu_\signal$ are calculated using asymptotic formulae and interpreted as limits on $\sigma(\tta) \times \mathcal{B}(\atautau)$.

All background and signal processes are affected by a set of associated systematic uncertainties. All the uncertainties are introduced in the fits as Gaussian nuisance parameters, with the exception of MC statistical uncertainties which are introduced via Poisson-constrained terms.

Theoretical uncertainties for all dominant backgrounds and signal samples originate from variations in the choice of PDFs, renormalisation scale and factorisation scale. Additional theoretical uncertainties are included for non-negligible processes involving a top-quark, related to modelling of initial- and final-state radiation, and from variations in the strong coupling constant and top-quark mass. For the dominant \ttbar, the sub-dominant \SingleTop, and the irreducible \tttautau processes, theoretical uncertainties due to the choice of generator and parton shower are also included, with comparisons performed between the nominal and the alternative samples listed in Section~\ref{sec:data-mc}.

The uncertainty in the combined 2015--2018 integrated luminosity is 0.83\% \cite{DAPR-2021-01}, obtained using the LUCID-2 detector \cite{LUCID2} for the primary luminosity measurements, complemented by measurements using the inner detector and calorimeters. An additional uncertainty for the MC pile-up reweighting is also considered.

The systematic variations due to the electron and muon reconstruction, identification and isolation efficiencies are evaluated by varying the correction factors for the selection efficiencies in the MC simulations by the corresponding uncertainties, estimated following the prescriptions in Refs.~\cite{EGAM-2021-02,MUON-2018-03}, studying the SM $Z\to\ell^+\ell^-$ and $W^{\pm} \to \mu^{\pm} \nu$ processes. Uncertainties in the lepton energy scale and resolution are included as variations in the lepton's momentum used in the event selection.
For jets, the energy scale and resolution uncertainties are derived following the procedure described in Ref.~\cite{JETM-2018-05}, and an uncertainty in the efficiency of the JVT requirement for pile-up suppression is also included~\cite{PERF-2014-03}. A set of $b$-tagging uncertainties is also included~\cite{FTAG-2019-07}.
The missing transverse energy uncertainties are estimated by propagating the uncertainties of the constituent objects included in the calculation, and additional uncertainties accounting for the scale and resolution of the soft term~\cite{JETM-2020-03}.

Boosted \dt objects include three types of systematic uncertainties. An uncertainty for the tau energy scale correction is estimated from variations of MC simulation conditions, and parameterised in the sub-jet number of tracks, the \dt system momentum and angular separation, and is applied on prompt boosted \dt objects in the analysis regions. Reconstruction and detector response uncertainties, calculated from variations in the detectors' geometry, and an identification scale-factor uncertainty calculated from data, are also applied on prompt boosted \dt objects in MC events~\cite{TAUP-2020-03}.

While the dominant contribution of events with fake boosted \dt objects from the \ttbar process is corrected through the \ttbar normalisation factor, other simulated backgrounds receive no such correction. To account for potential mismodelling of the fake boosted \dt objects in other backgrounds, an additional non-closure uncertainty is assigned to their background estimate. This uncertainty, binned in the di-$\tau$ invariant mass $m_{\tau\tau}$, is derived from the difference between data and the estimated MC background in a $Q_{\tau\tau}=1$ auxiliary region based on \CRttbar, but requiring exactly one or two jets. This additional jet requirement ensures that the region is orthogonal to all analysis regions defined in \cref{tab:event-selection:regions}. The modified selection reduces the $t\bar{t}$ purity by approximately 20\%. The MC event yield in this auxiliary region is obtained from a bootstrapped auxiliary background-only fit using the nominal configuration described above, excluding the fake-di-$\tau$ uncertainty under evaluation.

An uncertainty is assigned to the predicted fake-lepton background estimate described in Section~\ref{sec:stats:fakelep}. The modelling in the \VRFL is found to be satisfactory, with an overall non-closure in total yields of approximately 8\%. Despite this limited non-closure, a conservative uncertainty of 50\% is applied on the fake-lepton estimate to account for missing kinematic dependencies in the misidentification efficiencies, and their low-statistical precision.

A breakdown of the systematic uncertainties in the background estimates in the SR after a simultaneous \CRttbar+SR fit under the background-only hypothesis, grouped by the uncertainty type, is shown in \cref{tab:stats:systematics}. The overall uncertainty on the background prediction is dominated by statistical effects, while systematic uncertainties provide a smaller but non-negligible contribution. Among the sources of systematic uncertainty, the dominant contributions arise from theory uncertainties, in particular those affecting the \ttbar background, as well as MC statistical uncertainties. The uncertainties associated with the boosted \dt reconstruction and identification and with jet reconstruction constitute subleading contributions of comparable size. Within the latter, the largest effect originates from the jet energy resolution (JER), while the contribution from the jet energy scale (JES) remains below 1\%.

\begin{table}[ht]
\centering
\caption{Summary of the systematic uncertainties in the background expectation (in \%) in the  signal region, after a simultaneous fit to the control and signal regions under the background-only hypothesis. The individual uncertainties can be correlated and therefore do not necessarily add in quadrature to the total (stat.\ + syst.) background uncertainty. The \textit{Truth-Matched} boosted \dt uncertainty is applied to boosted \dt candidates whose two leading sub-jets are uniquely geometrically matched to the visible 4-momenta of two generator-level $\tauhad$. In all other cases, the {Fake} boosted \dt uncertainty is applied.}
\label{tab:stats:systematics}

% The next lines are included from the .//tables/statistical_analysis/SysTable_paper_styled.tex input file
\begin{tabular}{lc}
\toprule
\textbf{Uncertainty of region}                  &     SR [\%] \\
\midrule
Total (stat. + syst.) uncertainty               &        $12$ \\
\midrule
Theory total uncertainty                        &         $6$ \\
\quad $t\bar{t}$                                &         $6$ \\
\quad Z + jets                                  &       $< 1$ \\
\quad Diboson                                   &       $< 1$ \\
\quad Single top                                &       $< 1$ \\
\quad W + jets                                  &       $< 1$ \\
MC statistics                                   &       $2.7$ \\
DiTau total uncertainty                         &       $2.1$ \\
\quad Fake                                      &       $2.0$ \\
\quad Truth-matched                             &       $< 1$ \\
Jet total uncertainty                           &       $2.0$ \\
\quad JER                                       &       $1.9$ \\
\quad JES + in situ                             &       $< 1$ \\
\quad Flavour tagging                           &       $< 1$ \\
\quad JVT                                       &       $< 1$ \\
$t\bar{t}$ normalization ($\mu_{t\bar{t}}$)     &       $< 1$ \\
Electrons                                       &       $< 1$ \\
Muons                                           &       $< 1$ \\
Missing $E_T$                                   &       $< 1$ \\
Fake leptons bkg                                &       $< 1$ \\
Pileup reweighting and luminosity               &       $< 1$ \\
\bottomrule
\end{tabular}

% End of text imported from the .//tables/statistical_analysis/SysTable_paper_styled.tex input file

\end{table}

% End of text imported from the .//chapters/statistical_analysis.tex input file

%

%
\section{Results}
\label{sec:results}

\begin{figure}[ht]
\centering
\includegraphics[width=0.49\linewidth]{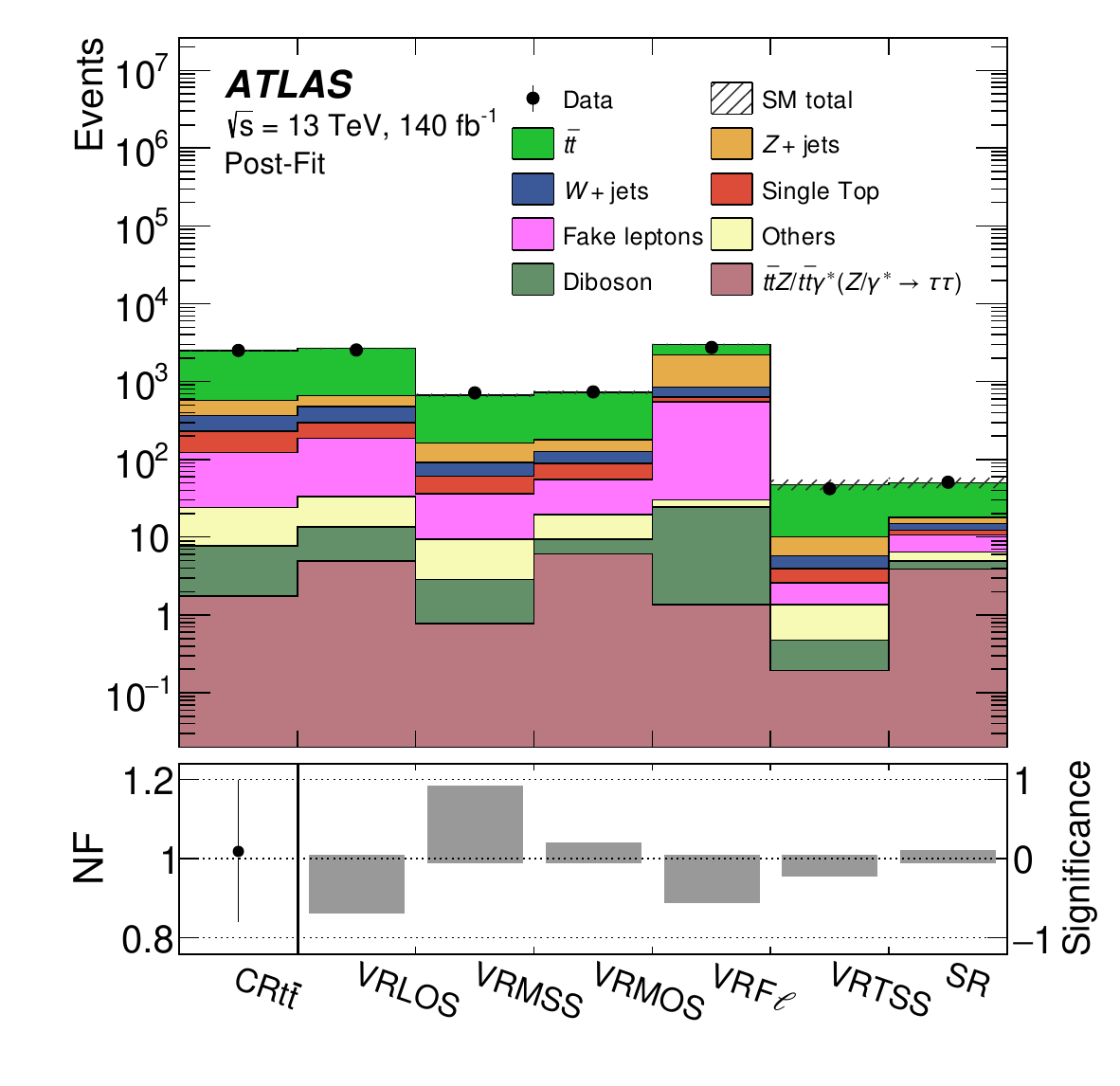}
\caption{Distribution of predicted and observed event yields in the analysis regions after a background-only fit in the \CRttbar. The lower panel shows the background normalisation factor (NF) for the \ttbar background estimated in the control region, and the significance, in standard deviations, of the difference between the observed and expected yields in the validation and signal regions, considering both the systematic and statistical uncertainties in the background expectation. The significance shown is calculated with the profile likelihood method from Ref.~\cite{Cowan:2010js}, adding a minus sign if the observed yield is below the prediction. Minor background processes, such as $t\bar{t}(Z \to \ell^+\ell^-/ q\bar{q})$, \ttW, \ttH, 3 top and 4 top productions are considered under `Others'.
}
\label{fig:results:pulls}
\end{figure}

% The next lines are included from the .//chapters/results.tex input file
A background-only fit was performed in the \CRttbar control region as discussed in \cref{sec:stats} to evaluate the SM background expectations in all regions of the analysis. The observed event yield in the \CRttbar region is used as a constraint in the fit to adjust the \ttbar background normalisation (assuming no signal contributes to this yield), obtaining a fitted normalisation factor value of $\mu_{\ttbar} = 1.02 \pm 0.18$. Excellent agreement between the estimated total background and the observed data events is seen in the signal region and in all the validation regions, as shown in \cref{fig:results:pulls}, within one standard deviation in all cases.

\begin{figure}[ht!]
\centering
\subfloat[]{\label{fig:result:distributions:ditaupt}\includegraphics[width=0.49\linewidth]{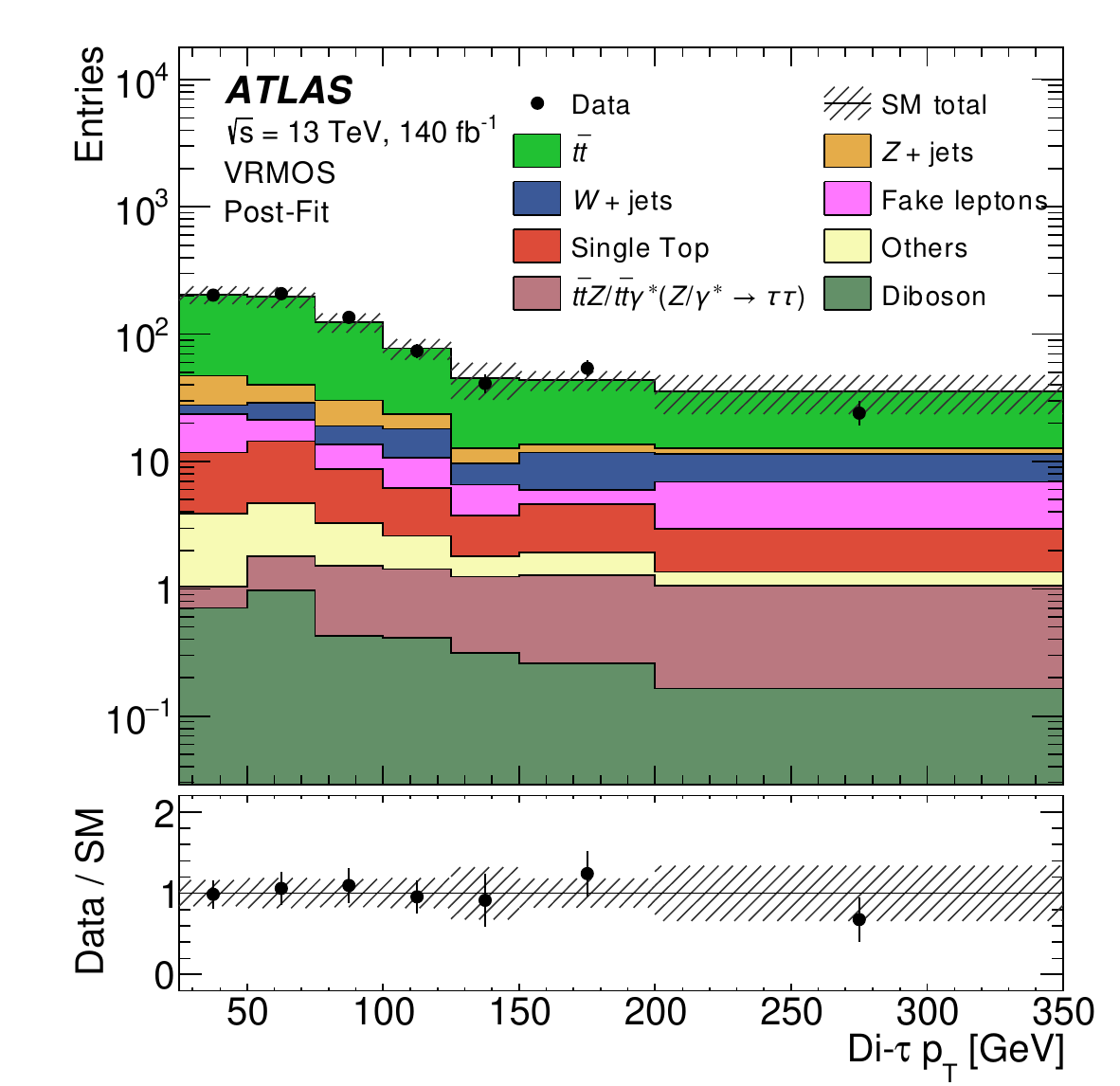}}
\subfloat[]{\label{fig:result:distributions:ditaum}\includegraphics[width=0.49\linewidth]{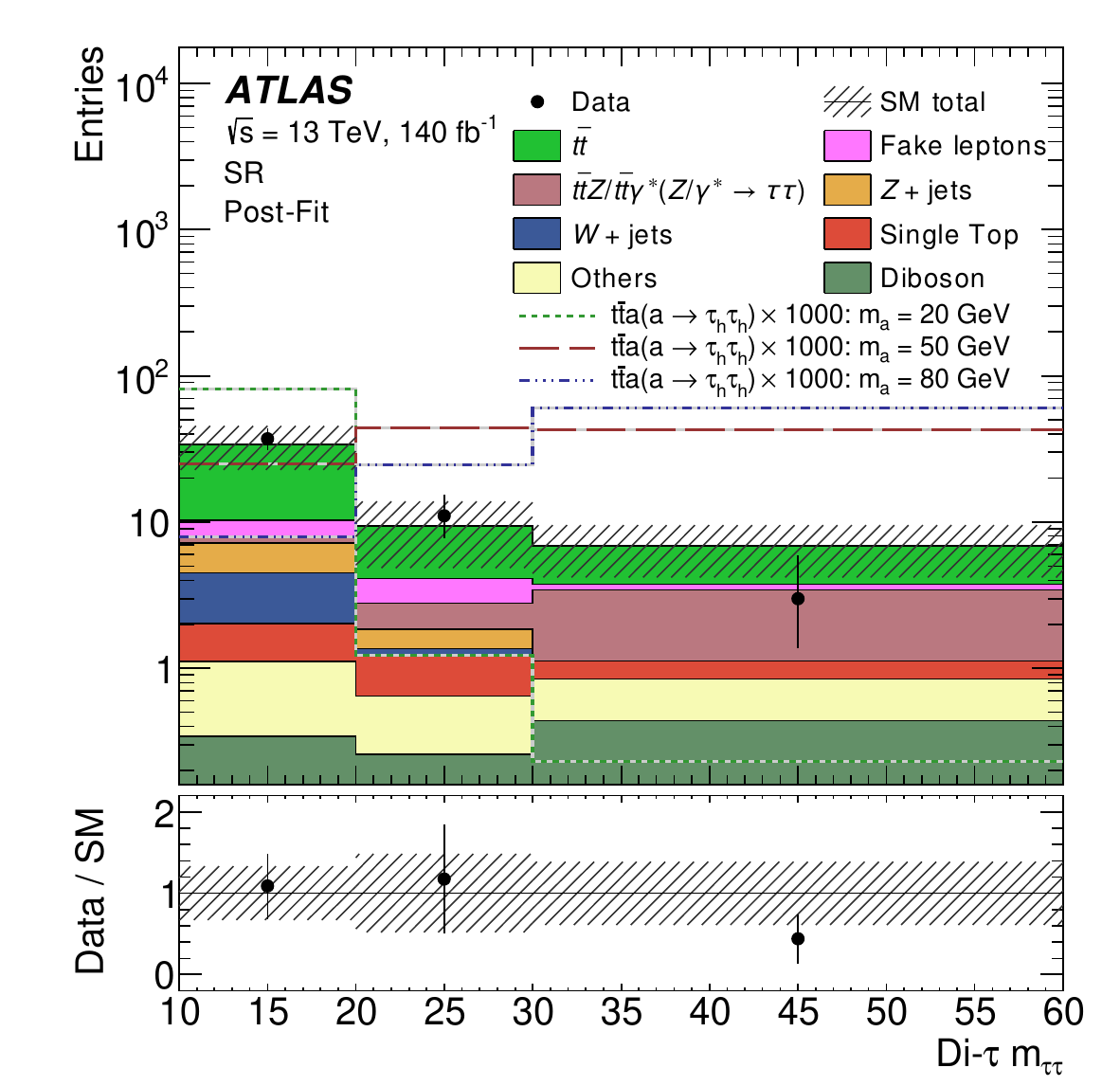}}
\caption{Observed distributions of the \protect\subref{fig:result:distributions:ditaupt} \dt \pT in the \VRMOS validation region and \protect\subref{fig:result:distributions:ditaum} \dt \mtt in the signal region, after a background-only fit in the \CRttbar. Statistical and systematic uncertainties are included in the background estimate uncertainty bands. Overflow events are included in the last bins. Minor background processes, such as $t\bar{t}(Z/\gamma^* \to \ell^+\ell^-/q\bar{q})$, \ttW, \ttH, 3 top and 4 top productions are considered under `Others'. The \SR distribution shows predictions for different masses of the 2HDM pseudoscalar signal studied in the search, scaled by a factor of $10^3$. The lower panels show the ratio of the data to the total SM background prediction.}
\label{fig:results:distributions}
\end{figure}

The predicted and observed \dt kinematic distributions agree throughout the kinematic phase-space of the analysis, exemplified by the \dt \pT and \mtautau distributions in the \VRMOS and \SR in \cref{fig:results:distributions}.

\cref{tab:results:yields} reports the expected background yields in the signal region, obtained from a simultaneous fit to the event yields in the \CRttbar and SR under the background-only hypothesis, in which the nuisance parameters and $\mu_{\ttbar}$ are fitted, together with the number of events observed in data. Results from the model-independent discovery fit are presented in \cref{tab:results:modelIndependent}, which provides the 95\%~CL upper limits on the number of BSM signal events ($S^{95}_{\text{obs}}$) and on the visible cross-section ($\langle A\varepsilon\sigma\rangle^{95}$). The corresponding expected limits, $S^{95}_{\text{exp}}$, are derived from an analogous hypothesis test performed on an Asimov dataset~\cite{Cowan:2010js} and are shown in the same table. The relatively large value of the model-independent upper limit on the signal yield, compared to the total background uncertainty of $\pm6$ events, arises from the increased uncertainty introduced in the discovery fit by the inclusion of an additional signal. The limited parameterisation of the likelihood model results in a reduced ability to constrain this extra component, thereby weakening the sensitivity of the fit and leading to a more conservative upper limit.

\begin{table}[t]
\centering
\small
\caption{Observed events and background estimates after the background-only fit in the \CRttbar and SR\@. The background uncertainties are both systematic and statistical. The individual uncertainties can be correlated and therefore do not necessarily add in quadrature to the total background uncertainty. Minor background processes, such as $t\bar{t}(Z/\gamma^* \to \ell^+\ell^-/q\bar{q})$, \ttW, \ttH, 3 top and 4 top productions are considered under `Others'.}

% The next lines are included from the .//tables/results/yields_cr_sr.tex input file
\begin{tabular}{l S[table-alignment-mode = format, table-format = 4.2]@{\(\,\pm\,\)}S[table-alignment-mode = format, table-format = 2.3] S@{\(\,\pm\,\)}S}
\toprule
& \multicolumn{2}{c}{$\text{CRt}\bar{\text{t}}$}           & \multicolumn{2}{c}{SR}                                     \\
\midrule
Observed events                                                                                 & \multicolumn{1}{S[table-alignment-mode = format, table-format = 4.2]@{\(\,\phantom{\pm}\,\)}}{2511}  &      & \multicolumn{1}{S@{\(\,\phantom{\pm}\,\)}}{51}    &        \\
\midrule
Expected SM events                                                                              & 2511                                              & 50   & 51                                                & 6      \\
\midrule
$\text{t}\bar{\text{t}}$                                                                        & 1935                                              & 98   & 33                                                & 7      \\
$Z$ + jets                                                                                      & 209                                               & 39   & \asymcol{3.2}{3.3}{3.2}{0pt}                               \\
$W$ + jets                                                                                      & 138                                               & 31   & 2.6                                               & 1.1    \\
Single top                                                                                      & 107                                               & 19   & 1.8                                               & 0.9    \\
Fake leptons                                                                                    & 98                                                & 49   & 4.1                                               & 2.1    \\
Others                                                                                          & 16.2                                              & 1.3  & 1.57                                              & 0.23   \\
Diboson                                                                                         & 6.0                                               & 1.4  & 1.04                                              & 0.28   \\
$\text{t}\bar{\text{t}}\text{Z}/\text{t}\bar{\text{t}}\gamma^*(\text{Z}/\gamma^*\to\tau\tau)$   & 1.75                                              & 0.24 & 3.9                                               & 1.0    \\
\bottomrule
\end{tabular}

% End of text imported from the .//tables/results/yields_cr_sr.tex input file

\label{tab:results:yields}
\end{table}

\begin{table}[t]
\centering
\small
\caption{For the SR, the observed model-independent 95\% CL upper limits on the visible cross-section ($\langle A\epsilon\sigma\rangle_{\text{obs}}^{95}$), the observed number of signal events ($S_{\text{obs}}^{95}$), and the number of signal events given the expected number of background events considering $\pm 1\sigma$ excursions on the expectation ($S_{\text{exp}}^{95}$) are shown. The last two columns indicate the number of events observed in data and the prediction of a background-only fit in the \CRttbar.}

% The next lines are included from the .//tables/results/ModelIndependent.tex input file
%
%
%
%
%
%
%
%
%
%

\begin{tabular}{lccccc}
\toprule
Region &
$\langle A\epsilon\sigma\rangle_{\mathrm{obs}}^{95}$ [fb] &
$S_{\mathrm{obs}}^{95}$ &
$S_{\mathrm{exp}}^{95}$ &
Observed events &
Expected SM events\\
\midrule
SR &
0.19 &
27.0 &
${26.6}^{+9.2}_{-6.8}$ &
51 &
50 $\pm13$\\
\bottomrule
\end{tabular}

% End of text imported from the .//tables/results/ModelIndependent.tex input file

\label{tab:results:modelIndependent}
\end{table}

\begin{figure}
\centering
\includegraphics[width=0.75\linewidth]{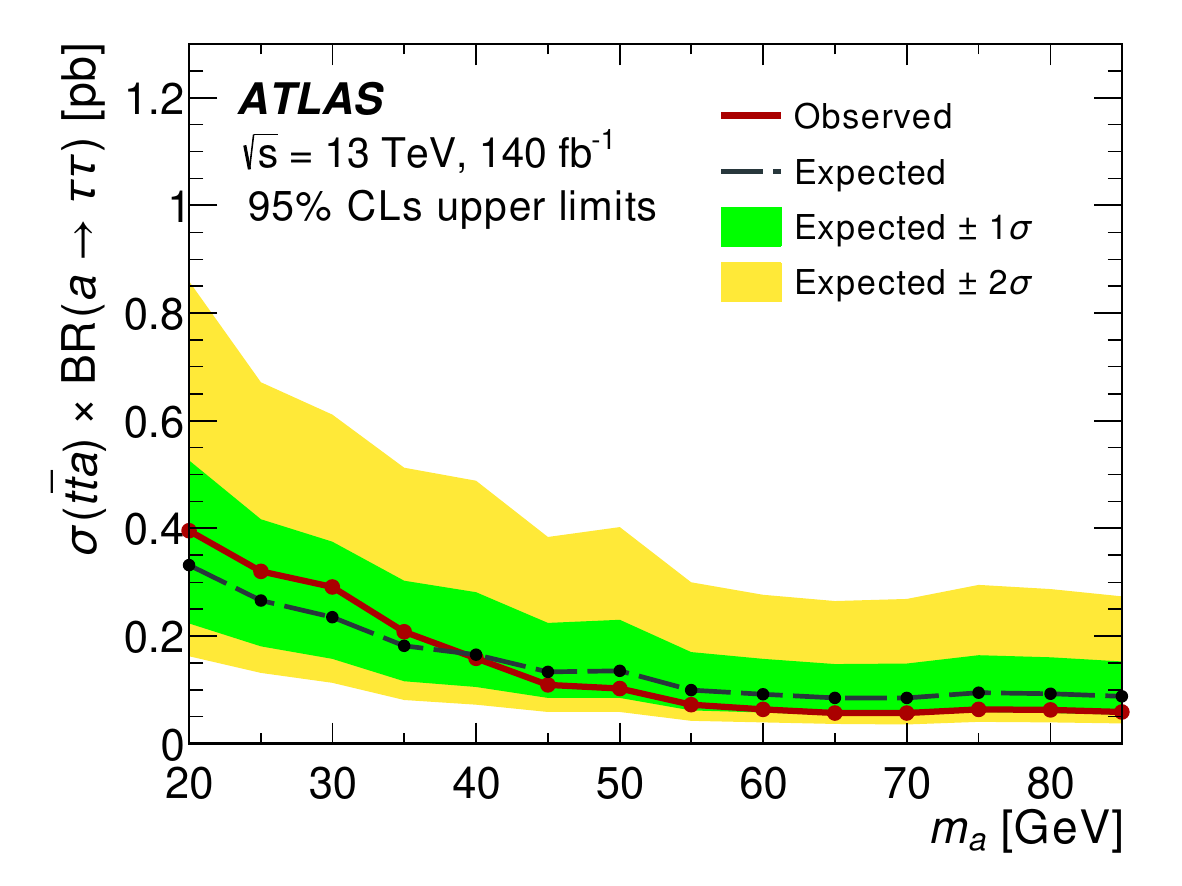}
\caption{Observed (solid line) and expected (dashed line) 95\% $\text{CL}s$ upper limits on a generic 2HDM \tta(\atautau) production cross-section times branching ratio, as a function of $m_a$. The shaded bands represent the $\pm1$ and $\pm2$ standard deviations around the expected limit.}
\label{fig:results:limits}
\end{figure}

\cref{fig:results:limits} presents the observed and expected 95\% $\text{CL}s$ upper limits for the 2HDM pseudoscalar\footnote{Scalar resonances with similar masses result in softer kinematic distributions for the \tau-leptons, relative to pseudoscalar resonances. The event and object selection used in this search do not include strict selections on the signal kinematics, therefore enabling reinterpretations of the results in terms of a scalar signal sharing the same \ttbar-associated production and final state.} signal presented in \cref{sec:data-mc}, as a function of $m_a$, accounting for the hadronic decay fraction of \tau-leptons. The limits range from approximately 0.4~pb for $m_a=\SI{20}{\GeV}$ to approximately 0.05~pb for $m_a=\SI{85}{\GeV}$. The decrease in sensitivity for lower $m_a$ values is driven by the acceptance$\times$efficiency factor, which decreases from approximately 0.14\% at $m_a=\SI{85}{\GeV}$ to 0.07\% at $m_a=\SI{20}{\GeV}$. This reduction is mainly due to average visible \tau-lepton transverse momenta decreasing as $m_a$ decreases, particularly for the sub-leading \tau-lepton, which limits the ability of the boosted \dt reconstruction algorithm to correctly capture such soft signatures in the sub-jets.

\FloatBarrier

% End of text imported from the .//chapters/results.tex input file

%

%
%
%

%
\section{Conclusion}
\label{sec:conclusion}

% The next lines are included from the .//chapters/conclusion.tex input file
A search for a low-mass Higgs-like resonance $a$, produced in association with a \ttbar pair, is performed using 140 fb$^{-1}$ of data recorded by the ATLAS experiment in proton–proton collisions at a centre-of-mass energy of $\sqrt{s}=\SI{13}{\TeV}$ at the LHC\@. The search considers decays of the resonance into a fully hadronic \dt final state, and targets a resonance mass range between \SI{20}{\GeV} and \SI{85}{\GeV}, where the \dt receives a significant Lorentz-boost. A dedicated algorithm is employed to accurately reconstruct the fully hadronic boosted \dt system and suppress SM background contributions, increasing the sensitivity at low masses compared to the traditional resolved \dt reconstruction.

A binned likelihood fit is performed to the SM background estimate and the observed data. The dominant \ttbar background normalisation is corrected in a control region, and the modelling of the background is verified in a set of validation regions spanning the phase-space between the control region and the signal region, observing a good agreement between data and SM predictions.

A good agreement between the SM background expectation and the measured data events is observed in the analysis signal region, and no significant excess is found. A model-independent discovery upper limit on the observed visible cross-section is established with a value of $0.19$ fb. Model-dependent exclusion upper limits on the cross-section times the branching ratio are calculated assuming a benchmark 2HDM model, and range from approximately \SI{0.4}{\pico\barn} for $m_a=\SI{20}{\GeV}$ to approximately \SI{0.05}{\pico\barn} for $m_a=\SI{85}{\GeV}$. This search provides the first ATLAS result in this boosted final state and complements recent ATLAS searches for top-quark associated $a$ production. The novel application of the fully hadronic boosted \dt reconstruction technique results in the strongest LHC limits for \ttatautau production in the hadronic \dt final state at \SI{20}{\GeV} at the time of publication.
%

% End of text imported from the .//chapters/conclusion.tex input file

%

%
%
%
%
%
\section*{Acknowledgements}
%

%

% The next lines are included from the .//acknowledgements/Acknowledgements.tex input file
%
%

%
%

We thank CERN for the very successful operation of the LHC and its injectors, as well as the support staff at
CERN and at our institutions worldwide without whom ATLAS could not be operated efficiently.

The crucial computing support from all WLCG partners is acknowledged gratefully, in particular from CERN, the ATLAS Tier-1 facilities at TRIUMF/SFU (Canada), NDGF (Denmark, Norway, Sweden), CC-IN2P3 (France), KIT/GridKA (Germany), INFN-CNAF (Italy), NL-T1 (Netherlands), PIC (Spain), RAL (UK) and BNL (USA), the Tier-2 facilities worldwide and large non-WLCG resource providers. Major contributors of computing resources are listed in Ref.~\cite{ATL-SOFT-PUB-2026-001}.

We gratefully acknowledge the support of ANPCyT, Argentina; YerPhI, Armenia; ARC, Australia; BMWFW and FWF, Austria; ANAS, Azerbaijan; CNPq and FAPESP, Brazil; NSERC, NRC and CFI, Canada; CERN; ANID, Chile; CAS, MOST, NSFC and FRFCU, China; Minciencias, Colombia; MEYS CR, Czech Republic; DNRF and DNSRC, Denmark; IN2P3-CNRS and CEA-DRF/IRFU, France; SRNSFG, Georgia; BMFTR, HGF and MPG, Germany; GSRI, Greece; RGC and Hong Kong SAR, China; ICHEP and Academy of Sciences and Humanities, Israel; INFN, Italy; MEXT and JSPS, Japan; CNRST, Morocco; NWO, Netherlands; RCN, Norway; MNiSW, Poland; FCT, Portugal; MNE/IFA, Romania; MSTDI, Serbia; MSSR, Slovakia; ARIS and MVZI, Slovenia; DSI/NRF, South Africa; MICIU/AEI, Spain; SRC and Wallenberg Foundation, Sweden; SERI, SNSF and Cantons of Bern and Geneva, Switzerland; NSTC, Taipei; TENMAK, T\"urkiye; STFC/UKRI, United Kingdom; DOE and NSF, United States of America.

Individual groups and members have received support from BCKDF, CANARIE, CRC and DRAC, Canada; CERN-CZ, FORTE and PRIMUS, Czech Republic; COST, ERC, ERDF, Horizon 2020 and Marie Sk{\l}odowska-Curie Actions, European Union; Investissements d'Avenir Labex, Investissements d'Avenir Idex and ANR, France; DFG and AvH Foundation, Germany; Herakleitos, Thales and Aristeia programmes co-financed by EU-ESF and the Greek NSRF, Greece; BSF-NSF and MINERVA, Israel; NCN and NAWA, Poland; La Caixa Banking Foundation, CERCA and AGAUR programs from Generalitat de Catalunya and PROMETEO and GenT Programmes Generalitat Valenciana, Spain; G\"{o}ran Gustafssons Stiftelse, Sweden; The Royal Society and Leverhulme Trust, United Kingdom; Eric and Wendy Schmidt Fund for Strategic Innovation, United States of America.

In addition, individual members wish to acknowledge support from Chile: Agencia Nacional de Investigaci\'on y Desarrollo (ANID FONDECYT reg. 1230987, FONDECYT 1230812, FONDECYT 1240864, Fondecyt 3240661, Fondecyt Regular 1240721); China: Fundamental Research Funds for the Central Universities (010-63263105), Chinese Ministry of Science and Technology (MOST-2023YFA1605700, MOST-2023YFA1609300), National Natural Science Foundation of China (NSFC 12275265, NSFC-W2543005); Czech Republic: Czech Science Foundation (GACR - 24-11373S), Ministry of Education Youth and Sports (ERC-CZ-LL2327, FORTE CZ.02.01.01/00/22\_008/0004632); EU: H2020 European Research Council (ERC - 101002463); European Union: European Research Council (BARD No. 101116429, ERC - 101219398, ERC 101089007), European Regional Development Fund (HE COFUND GA No.101081355, ERDF), Marie Sklodowska-Curie Actions (GAP-101168829); France: Agence Nationale de la Recherche (ANR-22-EDIR-0002, ANR-24-CE31-0504-01); Germany: Deutsche Forschungsgemeinschaft (DFG - 469666862); China: Research Grants Council (GRF); Italy: Istituto Nazionale di Fisica Nucleare (LHC-MIUR - 28003/2025), Ministero dell'Università e della Ricerca (NextGenEU I53D23000820006 M4C2.1.1, SOE2024\_0000023); Japan: Japan Society for the Promotion of Science (JSPS KAKENHI  JP25H0063, JSPS KAKENHI JP22H04944, JSPS KAKENHI JP24K23939, JSPS KAKENHI JP24KK0251, JSPS KAKENHI JP25H00650, JSPS KAKENHI JP25H01291, JSPS KAKENHI JP25K01011, JSPS KAKENHI JP25K01023, JSPS KAKENHI JP25KK0047); Poland: Polish National Science Centre (NCN 2021/42/E/ST2/00350, NCN OPUS 2023/51/B/ST2/02507, NCN OPUS nr 2022/47/B/ST2/03059, NCN UMO-2019/34/E/ST2/00393, UMO-2022/47/O/ST2/00148, UMO-2023/49/B/ST2/04085, UMO-2023/51/B/ST2/00920, UMO-2024/53/N/ST2/00869); Spain: Agència de Gestió d'Ajuts Universitaris i de Recerca. (AGAUR - 2023 BP 00141), Ministry of Science and Innovation (RYC2019-028510-I, RYC2020-030254-I, RYC2021-031273-I, RYC2022-038164-I), Ministerio de Ciencia, Innovación y Universidades/Agencia Estatal de Investigaci\'on (EU NextGenerationEU (PRTR-C17.I1), PID2022-142604OB-C22); Sweden: Carl Trygger Foundation (Carl Trygger Foundation CTS 22:2312), Swedish Research Council (Swedish Research Council 2023-04654, VR 2021-03651, VR 2022-03845, VR 2022-04683, VR 2023-03403, VR 2024-05451, VR 2025-05940), Knut and Alice Wallenberg Foundation (KAW 2023.0366); United Kingdom: The Binks Trust, Royal Society (NIF-R1-231091); United States of America: U.S. Department of Energy (ECA DE-AC02-76SF00515), John Templeton Foundation (John Templeton Foundation 63206), Neubauer Family Foundation.

%
%

% End of text imported from the .//acknowledgements/Acknowledgements.tex input file

%
\clearpage
\printbibliography
\clearpage
% ATLAS Collaboration author list
% Reference date of HMBS-2024-24 is 2026-06-15
% Author list last updated on date 30-JUL-26
% Data extracted on 30-Jul-2026 for paper reference HMBS-2024-24
% at 2:46am
 
\begin{flushleft}
\hypersetup{urlcolor=black}
{\Large The ATLAS Collaboration}

\bigskip

\AtlasOrcid[0000-0002-6665-4934]{G.~Aad}$^\textrm{\scriptsize 102}$,
\AtlasOrcid[0000-0001-7616-1554]{E.~Aakvaag}$^\textrm{\scriptsize 17}$,
\AtlasOrcid[0000-0002-5888-2734]{B.~Abbott}$^\textrm{\scriptsize 121}$,
\AtlasOrcid[0000-0002-0287-5869]{S.~Abdelhameed}$^\textrm{\scriptsize 83b}$,
\AtlasOrcid[0000-0002-1002-1652]{K.~Abeling}$^\textrm{\scriptsize 54}$,
\AtlasOrcid[0000-0001-5763-2760]{N.J.~Abicht}$^\textrm{\scriptsize 48}$,
\AtlasOrcid[0000-0002-8496-9294]{S.H.~Abidi}$^\textrm{\scriptsize 30}$,
\AtlasOrcid[0009-0003-6578-220X]{M.~Aboelela}$^\textrm{\scriptsize 44}$,
\AtlasOrcid[0000-0002-9987-2292]{A.~Aboulhorma}$^\textrm{\scriptsize 36e}$,
\AtlasOrcid[0000-0001-5329-6640]{H.~Abramowicz}$^\textrm{\scriptsize 154}$
% ...
\bigskip
\\

% ...
$^{17}$Department for Physics and Technology, University of Bergen, Bergen; Norway.\\
% ...
$^{30}$Physics Department, Brookhaven National Laboratory, Upton NY; United States of America.\\
% ...
$^{36}$$^{(a)}$Facult\'e des Sciences Ain Chock, Universit\'e Hassan II de Casablanca;$^{(b)}$Facult\'{e} des Sciences, Universit\'{e} Ibn-Tofail, K\'{e}nitra;$^{(c)}$Facult\'e des Sciences Semlalia, Universit\'e Cadi Ayyad, LPHEA-Marrakech;$^{(d)}$LPMR, Facult\'e des Sciences, Universit\'e Mohamed Premier, Oujda;$^{(e)}$Facult\'e des sciences, Universit\'e Mohammed V, Rabat;$^{(f)}$Institute of Applied Physics, Mohammed VI Polytechnic University, Ben Guerir; Morocco.\\
% ...
$^{44}$Physics Department, Southern Methodist University, Dallas TX; United States of America.\\
% ...
$^{48}$Fakult\"{a}t Physik, Technische Universit{\"a}t Dortmund, Dortmund; Germany.\\
% ...
$^{54}$II. Physikalisches Institut, Georg-August-Universit\"{a}t G\"ottingen, G\"ottingen; Germany.\\
% ...
$^{83}$$^{(a)}$Khalifa University of Science and Technology, Abu Dhabi;$^{(b)}$New York University Abu Dhabi, Abu Dhabi;$^{(c)}$United Arab Emirates University, Al Ain;$^{(d)}$University of Sharjah, Sharjah; United Arab Emirates.\\
% ...
$^{102}$CPPM, Aix-Marseille Universit\'e, CNRS/IN2P3, Marseille; France.\\
% ...
$^{121}$Homer L. Dodge Department of Physics and Astronomy, University of Oklahoma, Norman OK; United States of America.\\
% ...
$^{154}$Raymond and Beverly Sackler School of Physics and Astronomy, Tel Aviv University, Tel Aviv; Israel.\\
% ...
$^{a}$ Also at Affiliated with an institute formerly covered by a cooperation agreement with CERN.\\
$^{b}$ Also at An-Najah National University, Nablus; Palestine.\\
$^{c}$ Also at Borough of Manhattan Community College, City University of New York, New York NY; United States of America.\\
$^{d}$ Also at Center for Interdisciplinary Research and Innovation (CIRI-AUTH), Thessaloniki; Greece.\\
$^{e}$ Also at Centre of Physics of the Universities of Minho and Porto (CF-UM-UP); Portugal.\\
$^{f}$ Also at CERN, Geneva; Switzerland.\\
% ...

\end{flushleft}

% Created with Glance <Atlas.Glance@cern.ch>

%

%
%

%

%
%
%
%

\end{document}